\documentclass[twocolumn,showpacs,preprintnumbers,amsmath,amssymb,pra,aps,
    superscriptaddress,longbibliography]{revtex4-2}
\usepackage{graphicx}
\usepackage{multirow}
\usepackage{amsmath}
\usepackage{soul}
\usepackage{multirow}
\usepackage{verbatim}
\usepackage{bm}
\usepackage{outlines}
\usepackage{siunitx}
\usepackage[space]{grffile}
\usepackage[colorlinks=true, linkcolor=blue, citecolor=gray]{hyperref}
\usepackage[capitalize]{cleveref}
\usepackage{xr}
\usepackage{color}
\usepackage{tikz}
\usepackage{mathtools}
\usepackage{amssymb}
\usepackage{xcolor}
\usepackage{booktabs}
\usepackage[globalcitecopy]{bibunits}
\usepackage{cleveref}

\usepackage[usestackEOL]{stackengine}
\usepackage{scalerel}
\usepackage{graphicx,amsmath}

\usetikzlibrary{calc, positioning}

\newcommand{\TFD}{\ket{\mathrm{TFD}(\beta)}}
\newcommand{\cost}{\mathcal{C}}
\newcommand{\Uinter}{U_\textrm{inter}}
\newcommand{\Uintra}{U_\textrm{intra}}
\newcommand{\alphavec}{\vec{\alpha}}
\newcommand{\gammavec}{\vec{\gamma}}
\newcommand{\HA}{H_{\mathrm{A}}}

\newcommand{\HAB}{H_{\mathrm{BA}}}

\newcommand{\nA}{\ket{j}_\mathrm{A}}
\newcommand{\nB}{\ket{j}_\mathrm{B}}

\newcommand{\kB}{k_\mathrm{B}}

\newcommand{\groundState}{\Upsilon}
\newcommand{\optimizedState}{\Psi}

\newcommand{\ket}[1]{\left\lvert #1 \right\rangle}
\newcommand{\bra}[1]{\left\langle #1 \right\rvert}

\newcommand{\rhoGibbs}{\rho_{\mathrm{Gibbs}}}
\newcommand{\rhoexp}{\rho_{\mathrm{Exp}}}
\newcommand{\rhotwoqutrit}{\rho_{\mathrm{2Qutrit}}}

\newcommand{\Rx}{R_X}
\newcommand{\Ry}{R_Y}
\newcommand{\Rz}{R_Z}
\newcommand{\Rmw}{R_{\mathrm{XY}}}

\newcommand{\XAone}{X_{\mathrm{A1}}}
\newcommand{\XAtwo}{X_{\mathrm{A2}}}
\newcommand{\ZAone}{Z_{\mathrm{A1}}}
\newcommand{\ZAtwo}{Z_{\mathrm{A2}}}

\newcommand{\XBone}{X_{\mathrm{B1}}}
\newcommand{\XBtwo}{X_{\mathrm{B2}}}
\newcommand{\ZBone}{Z_{\mathrm{B1}}}
\newcommand{\ZBtwo}{Z_{\mathrm{B2}}}

\newcommand{\XA}{X_{\mathrm{A}}}
\newcommand{\XB}{X_{\mathrm{B}}}
\newcommand{\XXA}{X\! X_{\mathrm{A}}}
\newcommand{\XXB}{X\! X_{\mathrm{B}}}
\newcommand{\XXAB}{X\! X_{\mathrm{BA}}}

\newcommand{\YA}{Y_{\mathrm{A}}}
\newcommand{\YB}{Y_{\mathrm{B}}}
\newcommand{\YYA}{Y\! Y_{\mathrm{A}}}
\newcommand{\YYB}{Y\! Y_{\mathrm{B}}}
\newcommand{\YYAB}{Y\! Y_{\mathrm{BA}}}

\newcommand{\ZA}{Z_{\mathrm{A}}}
\newcommand{\ZB}{Z_{\mathrm{B}}}
\newcommand{\ZZA}{Z\! Z_{\mathrm{A}}}
\newcommand{\ZZB}{Z\! Z_{\mathrm{B}}}
\newcommand{\ZZAB}{Z\! Z_{\mathrm{BA}}}

\newcommand{\II}{I\! I}

\newcommand{\XX}{X\! X}
\newcommand{\ZZ}{Z\! Z}

\newcommand{\QAone}{\mathrm{A}_1}
\newcommand{\QAtwo}{\mathrm{A}_2}
\newcommand{\QBone}{\mathrm{B}_1}
\newcommand{\QBtwo}{\mathrm{B}_2}
\newcommand{\Tone}{T_{1}}
\newcommand{\Ttwoecho}{T_2^{\textrm{echo}}}

\newcommand{\invTemp}{\beta}

\newcommand{\us}{\mu\mathrm{s}}
\newcommand{\GHz}{\mathrm{GHz}}
\newcommand{\K}{\mathrm{K}}
\newcommand{\mK}{\mathrm{mK}}
\newcommand{\dB}{\mathrm{dB}}
\newcommand{\ns}{\mathrm{ns}}
\newcommand{\degrees}{^{\circ}}
\newcommand{\nm}{\mathrm{nm}}

\newcommand{\tr}{\text{Tr}}
\newcommand{\expect}[3]{\left\langle #1 \middle\rvert #2 \middle\lvert #3 \right\rangle}

\begin{document}
\title{Variational preparation of finite-temperature states on a quantum computer}

%%%%% BEGIN OF AUTHORS %%%%%%
\author{R.~Sagastizabal}
\thanks{These authors contributed equally to this work.}
\affiliation{QuTech, Delft University of Technology, P.O. Box 5046, 2600 GA Delft, The Netherlands}
\affiliation{Kavli Institute of Nanoscience, Delft University of Technology, P.O. Box 5046, 2600 GA Delft, The Netherlands}

\author{S.~P.~Premaratne}
\thanks{These authors contributed equally to this work.}
\affiliation{Intel Labs, Intel Corporation, Hillsboro, Oregon 97124, USA}

\author{B.~A.~Klaver}
\affiliation{QuTech, Delft University of Technology, P.O. Box 5046, 2600 GA Delft, The Netherlands}
\affiliation{Kavli Institute of Nanoscience, Delft University of Technology, P.O. Box 5046, 2600 GA Delft, The Netherlands}

\author{M.~A.~Rol}
\affiliation{QuTech, Delft University of Technology, P.O. Box 5046, 2600 GA Delft, The Netherlands}
\affiliation{Kavli Institute of Nanoscience, Delft University of Technology, P.O. Box 5046, 2600 GA Delft, The Netherlands}

\author{V.~Neg\^{\i}rneac}
\affiliation{QuTech, Delft University of Technology, P.O. Box 5046, 2600 GA Delft, The Netherlands}
\affiliation{Instituto Superior T\'{e}cnico, Lisbon, Portugal}

\author{M.~Moreira}
\affiliation{QuTech, Delft University of Technology, P.O. Box 5046, 2600 GA Delft, The Netherlands}
\affiliation{Kavli Institute of Nanoscience, Delft University of Technology, P.O. Box 5046, 2600 GA Delft, The Netherlands}

\author{X.~Zou}
\affiliation{Intel Labs, Intel Corporation, Hillsboro, Oregon 97124, USA}

\author{S.~Johri}
\thanks{Present address: IonQ, College Park, MD 20740, USA.}
\affiliation{Intel Labs, Intel Corporation, Hillsboro, Oregon 97124, USA}

\author{N.~Muthusubramanian}
\affiliation{QuTech, Delft University of Technology, P.O. Box 5046, 2600 GA Delft, The Netherlands}
\affiliation{Kavli Institute of Nanoscience, Delft University of Technology, P.O. Box 5046, 2600 GA Delft, The Netherlands}

\author{M.~Beekman}
\affiliation{Netherlands Organisation for Applied Scientific Research (TNO), P.O. Box 155, 2600 AD Delft, The Netherlands}
\affiliation{QuTech, Delft University of Technology, P.O. Box 5046, 2600 GA Delft, The Netherlands}
\author{C.~Zachariadis}
\affiliation{QuTech, Delft University of Technology, P.O. Box 5046, 2600 GA Delft, The Netherlands}
\affiliation{Kavli Institute of Nanoscience, Delft University of Technology, P.O. Box 5046, 2600 GA Delft, The Netherlands}

\author{V.~P.~Ostroukh}
\affiliation{QuTech, Delft University of Technology, P.O. Box 5046, 2600 GA Delft, The Netherlands}
\affiliation{Kavli Institute of Nanoscience, Delft University of Technology, P.O. Box 5046, 2600 GA Delft, The Netherlands}

\author{N.~Haider}
\affiliation{Netherlands Organisation for Applied Scientific Research (TNO), P.O. Box 155, 2600 AD Delft, The Netherlands}
\affiliation{QuTech, Delft University of Technology, P.O. Box 5046, 2600 GA Delft, The Netherlands}
\author{A.~Bruno}
\affiliation{QuTech, Delft University of Technology, P.O. Box 5046, 2600 GA Delft, The Netherlands}
\affiliation{Kavli Institute of Nanoscience, Delft University of Technology, P.O. Box 5046, 2600 GA Delft, The Netherlands}

\author{A.~Y.~Matsuura}
\affiliation{Intel Labs, Intel Corporation, Hillsboro, Oregon 97124, USA}

\author{L.~DiCarlo}
\affiliation{QuTech, Delft University of Technology, P.O. Box 5046, 2600 GA Delft, The Netherlands}
\affiliation{Kavli Institute of Nanoscience, Delft University of Technology, P.O. Box 5046, 2600 GA Delft, The Netherlands} 
%%%%% END OF AUTHORS %%%%%%%%

\date{\today}

\begin{abstract}
The preparation of thermal equilibrium states is important for the simulation of condensed-matter and cosmology systems using a quantum computer.
We present a method to prepare such mixed states with unitary operators, and demonstrate this technique experimentally using a gate-based quantum processor.
Our method targets the generation of thermofield double states using a hybrid quantum-classical variational approach motivated by quantum-approximate optimization algorithms, without prior calculation of optimal variational parameters by numerical simulation.
The fidelity of generated states to the thermal-equilibrium state smoothly varies from $99$ to $75\%$ between infinite and near-zero simulated temperature, in
quantitative agreement with numerical simulations of the noisy quantum processor with error parameters drawn from experiment.
\end{abstract}
\maketitle

\begin{bibunit}[apsrev4-2]

\section*{Introduction}
The potential for quantum computers to simulate other quantum mechanical systems is well known~\cite{Feynman82}, and the ability to represent the dynamical evolution of quantum many-body systems has been demonstrated~\cite{Lloyd96}. However, the accuracy of these simulations depends on efficient initial state preparation within the quantum computer. Much progress has been made on the efficient preparation of non-trivial quantum states, including spin-squeezed states~\cite{Hosten16} and entangled cat states~\cite{Vlastakis13}. Studying phenomena like high-temperature superconductivity~\cite{Lee06} requires preparation of thermal equilibrium states, or Gibbs states. Producing mixed states with unitary quantum operations is not straightforward, and has only recently begun to be explored~\cite{Zhu19, Chowdhury20}. In this work, we demonstrate the use of a variational quantum-classical algorithm to realize Gibbs states using (ideally unitary) gate control on a transmon quantum processor.

Our approach is mediated by the generation of thermofield double (TFD) states, which are pure states sharing entanglement between two identical quantum systems with the characteristic that when one of the systems is considered independently (by tracing over the other), the result is a mixed state representing equilibrium at a specific temperature. TFD states are of interest not only in condensed matter physics but also for the study of black holes~\cite{Israel76, Maldacena03} and traversable wormholes~\cite{Maldacena17, Gao17}. We use a variational protocol~\cite{Wu19} motivated by quantum-approximate optimization algorithms (QAOA) that relies on alternation of unitary intra- and inter-system operations to control the effective temperature, eliminating the need for a large external heat bath. Recently, verification of TFD state preparation was demonstrated on a trapped-ion quantum computer~\cite{Zhu19}. Our work experimentally demonstrates the first generation of finite-temperature states in a superconducting quantum computer by variational preparation of TFD states in a hybrid quantum-classical manner.

\section*{Results}
\subsection*{Theory}
Consider a quantum system described by Hamiltonian $H$ with eigenstates $\ket{j}$ and corresponding eigenenergies $E_j$:
\[
H \ket{j} = E_{j} \ket{j}.
\]
The Gibbs state $\rhoGibbs$ of the system is
\[
\rhoGibbs(\invTemp)= \frac{1}{Z} \sum_{j} \exp \left(-\invTemp E_j\right) \ket{j}\bra{j},
\]
where $\invTemp=1/\kB T$ is the inverse temperature, $\kB$ is the Boltzmann constant, and
\[
Z=\sum_j \exp \left(-\invTemp E_j\right)
\]
is the partition function.
Except in the limit $\beta \rightarrow \infty$, the Gibbs state is a mixed state and thus impossible to generate strictly through unitary evolution.
To circumvent this, we define the TFD state~\cite{Wu19} on two identical systems $\mathrm{A}$ and $\mathrm{B}$ as
\[
\ket{\textrm{TFD}(\invTemp)} = \frac{1}{\sqrt{Z}} \sum_{j} \exp \left( \frac{-\invTemp E_j}{2} \right) \nB\! \nA.
\]
Tracing out either system yields the desired Gibbs state in the other.

\begin{figure}
\begin{center}
\includegraphics[width=\columnwidth]{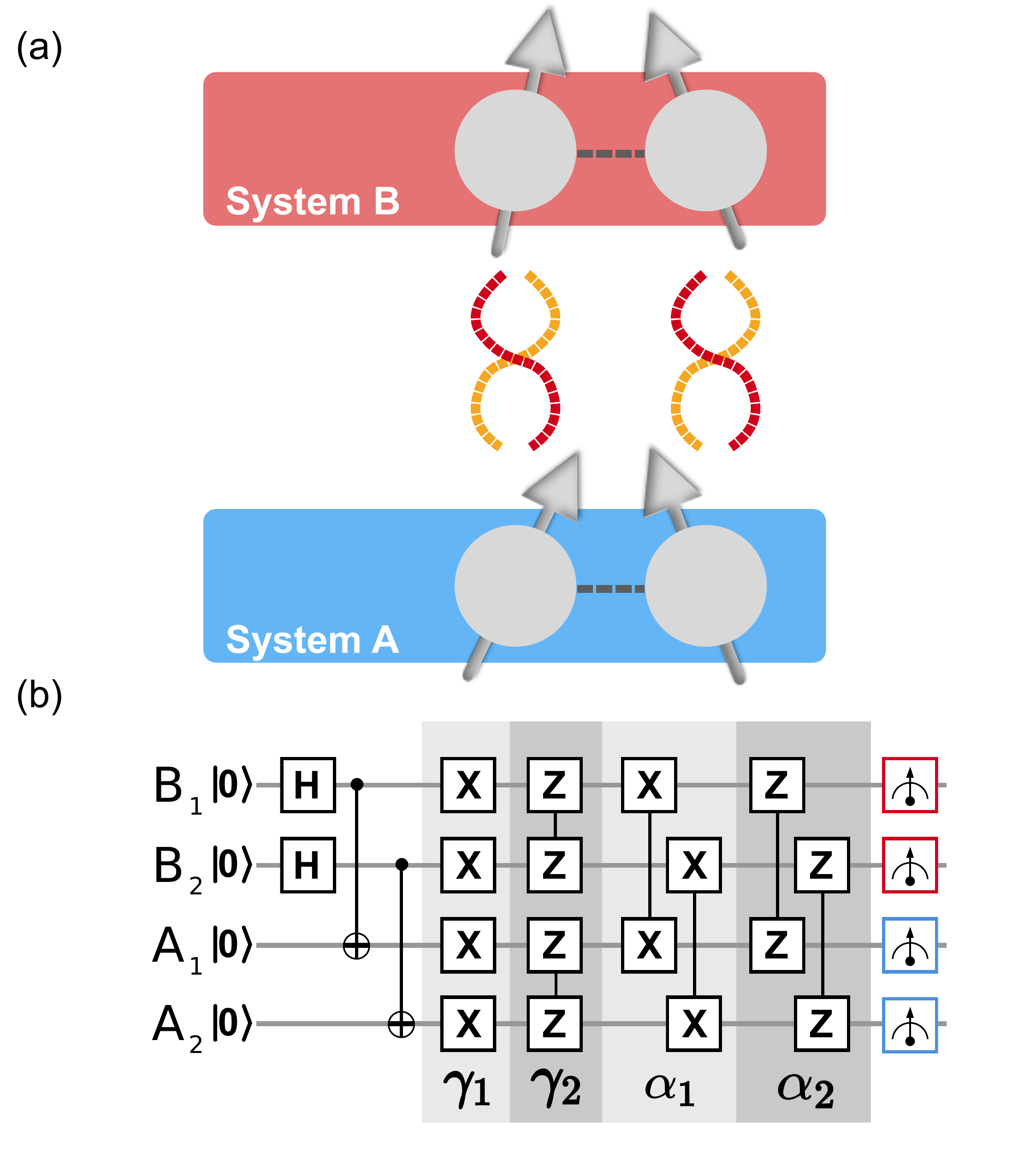}
\end{center}
\caption{
\label{fig:Label_FIG1}
\textbf{Principle and generation of Thermofield Double state.}
(a) Two identical systems A and B are variationally prepared in an ideally pure, entangled joint state such that tracing out one system yields the Gibbs state on the other. (b) Corresponding qubits in the two systems are first pairwise entangled to produce the $\invTemp = 0$ TFD state. Next, intra- and inter-system Hamiltonians are applied with optimized variational angles $\left(\alphavec,\gammavec\right)$ to approximate the TFD state corresponding to the desired temperature.
}
\end{figure}

To prepare the TFD states, we follow the variational protocol proposed by~\citet{Wu19} and consider two systems each of size $n$.
In the first step of the procedure, the TFD state at $\invTemp=0$ is generated by creating Bell pairs
$\ket{\Phi_i^+} = \left(\ket{0}_{\mathrm{B}i} \ket{0}_{\mathrm{A}i} + \ket{1}_{\mathrm{B}i} \ket{1}_{\mathrm{A}i}\right)/\sqrt{2}$
between corresponding qubits $i$ in the two systems. Tracing out either system yields a maximally mixed state on the other, and vice versa.
The next steps to create the TFD state at finite temperature depend on the relevant Hamiltonian.
Here, we choose the transverse field Ising model in a one-dimensional chain of $n$ spins~\cite{Ho19}, with $n = 2$ [\cref{fig:Label_FIG1}(a)].
We map spin up (down) to the computational state $\ket{0}$ $(\ket{1})$ of the corresponding transmon.
The Hamiltonian describing system A is
\[
\HA = \ZZA + g \XA,
\]
where $\ZZA=\ZAtwo\ZAone$, $\XA=\XAtwo+\XAone$, and $g$ is proportional to the transverse magnetic field. The Hamiltonian for system B is the same.
We focus on $g=1$, where a phase transition is expected in the transverse field Ising model at large $n$~\cite{Bonfim19}.
We use a QAOA-motivated variational ansatz~\cite{Wu19, Hadfield19}, where intra-system evolution is interleaved with a Hamiltonian enforcing interaction between the systems:
\[
\HAB= \XXAB + \ZZAB,
\]
where $\XXAB=\XBtwo \XAtwo + \XBone \XAone$, and analogously for $\ZZAB$.
For single-step state generation, the unitary operation describing the TFD protocol is
\[
U \left(\alphavec, \gammavec \right)= \Uinter \left(\alphavec\right) \Uintra \left(\gammavec\right),
\]
where
\[
\begin{split}
\Uintra(\gammavec) =& \exp \left(-i\gamma_2 \left( \ZZB + \ZZA \right)/2\right) \\
                                & \times \exp \left(-i \gamma_1 \left( \XB + \XA \right)/2\right), \\
\Uinter(\alphavec) =& \exp \left(-i \alpha_2 \ZZAB/2\right) \exp \left(-i \alpha_1 \XXAB/2\right).
\end{split}
\]
The variational parameters $\gammavec=\left(\gamma_1, \gamma_2\right)$, $\alphavec=\left(\alpha_1, \alpha_2\right)$ are optimized by the hybrid classical-quantum algorithm to generate states closest to the ideal TFD states. A single step of intra- and inter-system interaction ideally produces the state $\ket{\psi (\alphavec, \gammavec)} = U  \left(\alphavec, \gammavec \right) \left( \ket{\Phi^{+}_{2}}\otimes \ket{\Phi^{+}_{1}} \right)$~\cite{Premaratne20}.

The variational algorithm extracts the cost function after each state preparation. We engineer a cost function $\cost$ to be minimized when the generated state is closest to an ideal TFD state~\cite{Premaratne20}.
This cost function is
\[
\begin{split}
C(\invTemp) &= \langle \XA \rangle + \langle \XB \rangle  + 1.57 \left( \langle\ZZA\rangle + \langle\ZZB\rangle \right) \\
&\hspace{8em} - \invTemp^{-1.57} (\langle \XXAB \rangle + \langle \ZZAB \rangle ).
\end{split}
\]
We compare the performance of this engineered cost function $\cost_{1.57}$ to that of the non-optimized cost function $\cost_{1.00}$, using the reduction of infidelity to the Gibbs state as the ultimate metric of success (see~\cite{SOM_QAOA}). The engineered cost function achieves an average improvement of $54\%$ across the $\invTemp$ range covered ($[10^{-2},10^{2}]$ in units of $1/g$), as well as a maximum improvement of up to $98\%$ for intermediate temperatures ($\invTemp \sim 1$). Our choice of the class of cost functions to optimize lets us trade off a slight decrease in low-temperature performance with a significant increase in performance at intermediate temperatures. See~\cite{Premaratne20} for further details on the theory.

The quantum portion of the algorithm prepares the state according to a given set of angles $\left(\alphavec, \gammavec \right)$, performs the measurements, and returns these values to the classical portion. The classical portion then evaluates the cost function according to the returned measurements, performs classical optimization, generates and returns the next set of variational angles to evaluate on the quantum portion.

\subsection*{Experiment}

\begin{figure}
\begin{center}
\includegraphics[width=\columnwidth]{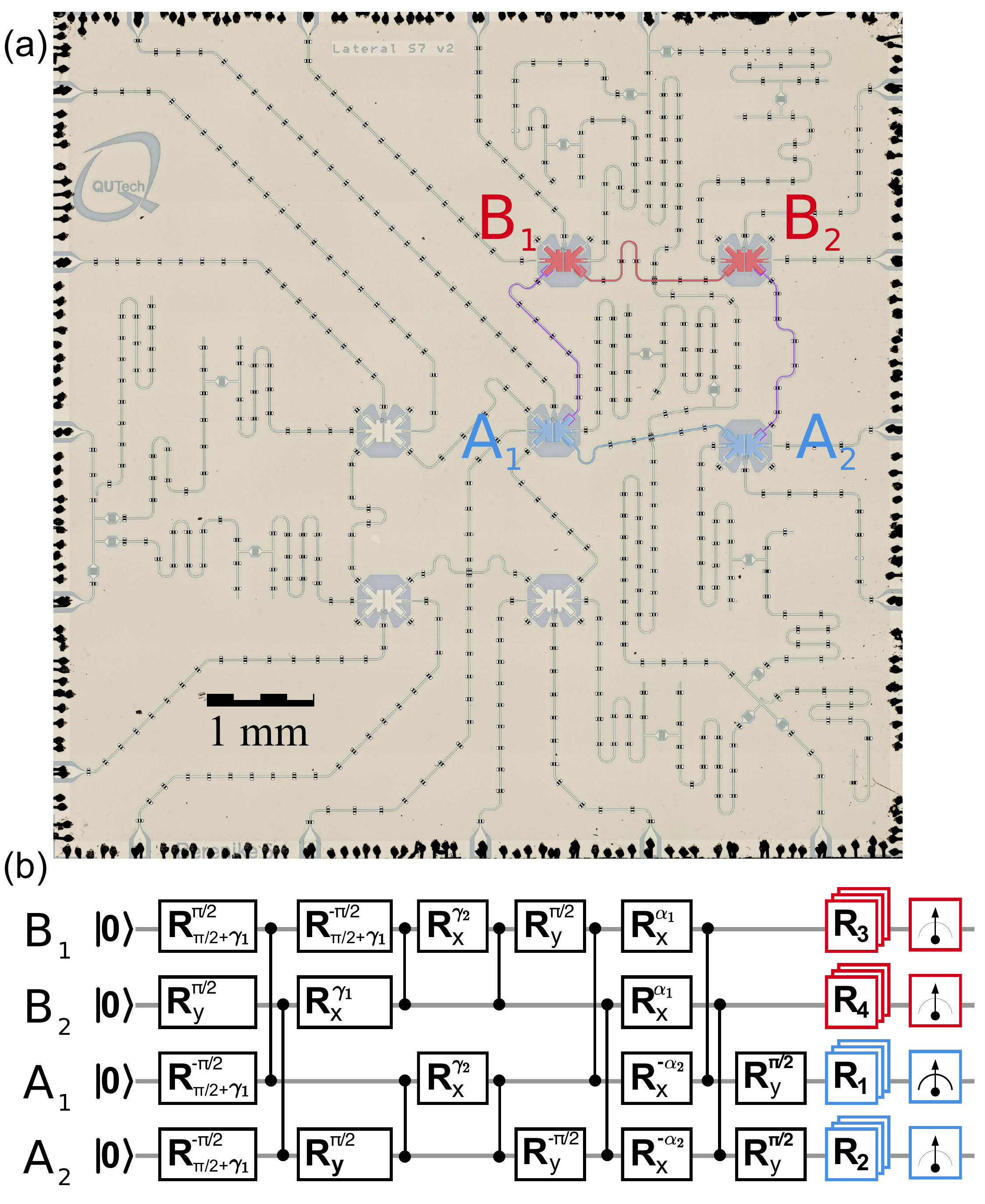}
\end{center}
\caption{
\label{fig:Label_FIG2}
\textbf{Device and optimized quantum circuit.}
(a) Optical image of the transmon processor used in this experiment, with false color highlighting the four transmons employed and the dedicated bus resonators providing their nearest-neighbor coupling.
(b) Optimized circuit equivalent to that in \cref{fig:Label_FIG1}(b) and conforming to the native gate set in our architecture.  All variational parameters are mapped onto rotation axes and angles of single-qubit gates.
Tomographic pre-rotations $R_1$-$R_4$ are added to reconstruct the terms in the cost function $\cost$ and to perform two-qubit state tomography of each system following optimization.
}
\end{figure}

We implement the algorithm using four of seven transmons in a monolithic quantum processor [\cref{fig:Label_FIG2}(a)]. The four transmons (labelled $\QAone$, $\QAtwo$, $\QBone$, and $\QBtwo$) have square connectivity provided by coupling bus resonators, and are thus ideally suited for implementing the circuit in \cref{fig:Label_FIG1}(b). Each transmon has a microwave-drive line for single-qubit gating, a flux-bias line for two-qubit controlled-$Z$ (CZ) gates, and a dispersively coupled resonator with dedicated Purcell filter~\cite{Heinsoo18,Bultink20}. The four transmons can be simultaneously and independently read out by frequency multiplexing, using the common feedline connecting to all Purcell filters. All transmons are biased to their flux-symmetry point (i.e., sweetspot~\cite{Schreier08}) using static flux bias to counter residual offsets. Device details and a summary of measured transmon parameters are provided in~\cite{SOM_QAOA}.

In order to realize the theoretical circuit in \cref{fig:Label_FIG1}(b), we first map it to the optimized depth-13 equivalent circuit shown in \cref{fig:Label_FIG2}(b), which conforms to the native gate set in our control architecture. This gate set consists of arbitrary single-qubit rotations about any equatorial axis of the Bloch sphere, and CZ gates between nearest-neighbor transmons. Conveniently, all variational angles are mapped to either the axis or angle of single-qubit rotations. Further details on the compilation steps are reported in the Methods section and~\cite{SOM_QAOA}.  Bases pre-rotations are added at the end of the circuit to first extract all the terms in the cost function $\cost$ and finally to perform two-qubit state tomography of each system.

\begin{figure}
\begin{center}
\includegraphics[trim={0 0.97cm 0 0},clip,width=0.5\textwidth]{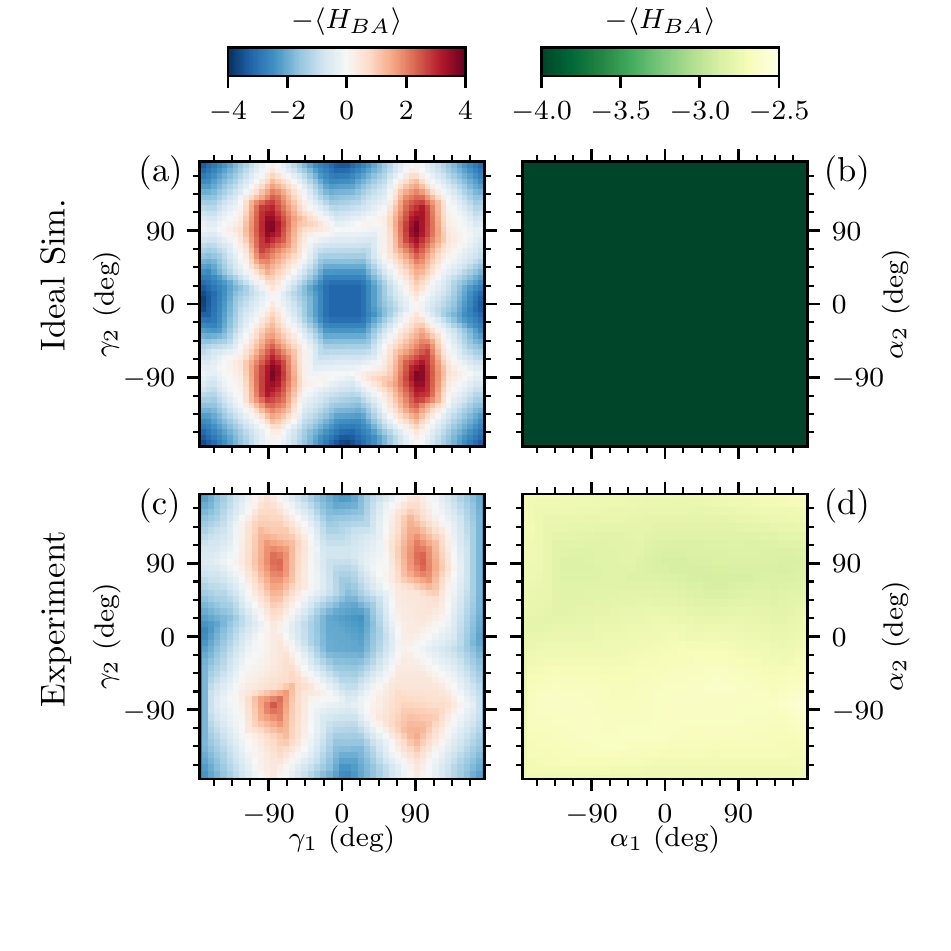}
\end{center}
\caption{
\label{fig:Label_FIG3}
\textbf{Landscape of the cost function for infinite temperature.}
Panels (a) and (c): Landscape cuts obtained in noiseless simulation and in experiment, respectively, when varying control angles $\gammavec$ while keeping $\alphavec=0$ (the ideal solution value).
Panels (b) and (d): Corresponding cuts obtained when varying control angles $\alphavec$ while keeping $\gammavec=0$ (the ideal solution value). See text for details. These landscape cuts are sampled at $100$ points and interpolated using the Python package \textit{adaptive}~\cite{Nijholt19}.
}
\end{figure}

Prior to implementing any variational optimizer, it is helpful to build a basic understanding of the cost-function landscape.
To this end, we investigate the cost function $\cost$  at $\invTemp=0$ using two-dimensional cuts: we sweep $\gammavec$ while keeping $\alphavec=0$ to study the effect of $\Uintra$  and vice versa to study the effect of $\Uinter$.
Note that owing to the $\invTemp^{-1.57}$ divergence, the cost function reduces to $-\langle \HAB \rangle$ in the $\invTemp=0$ limit.
Consider first the landscape for an ideal quantum processor, which is possible to compute for our system size.
The $\gammavec$ landscape at $\alphavec=0$ is $\pi$-periodic in both directions due to the invariance of $\ket{\textrm{TFD}(\invTemp=0)}$ under
bit-flip $(X)$ and phase-flip $(Z)$ operations on all qubits. The cost function is minimized to -4 at even multiples of $\pi/2$ on $\gamma_1$ and $\gamma_2$: $\ket{\textrm{TFD}(\invTemp=0)}$ is a simultaneous eigenstate of
$\XXAB$ and $\ZZAB$ with eigenvalue +2 due to the symmetry of the constituting Bell states $\ket{\Phi_i^+}$. In turn, the cost function is maximized to +4 at odd multiples of $\pi/2$, at which the $\ket{\Phi_i^+}$ are transformed to
singlets $\ket{\Psi_i^-} = \left(\ket{0}_{\mathrm{B}i} \ket{1}_{\mathrm{A}i} - \ket{1}_{\mathrm{B}i} \ket{0}_{\mathrm{A}i}\right)/\sqrt{2}$.
The $\alphavec$ landscape at $\gammavec=0$ is constant, reflecting that $\ket{\textrm{TFD}(\invTemp=0)}$ is a simultaneous eigenstate of $\XXAB$ and $\ZZAB$ and thus also of any exponentiation of these operators.
The corresponding experimental landscapes show qualitatively similar behavior. The $\gammavec$ landscape clearly shows the $\pi$ periodicity with respect to both angles, albeit with reduced contrast. The $\alphavec$ landscape is not strictly constant, showing weak structure particularly with respect to $\alpha_2$. These experimental deviations reflect underlying errors in our noisy intermediate-scale quantum (NISQ) processor, which include transmon decoherence, residual $ZZ$ coupling at the bias point, and leakage during CZ gates. We discuss these error sources in detail further below.

\begin{figure}
\begin{center}
\includegraphics[width=0.5\textwidth]{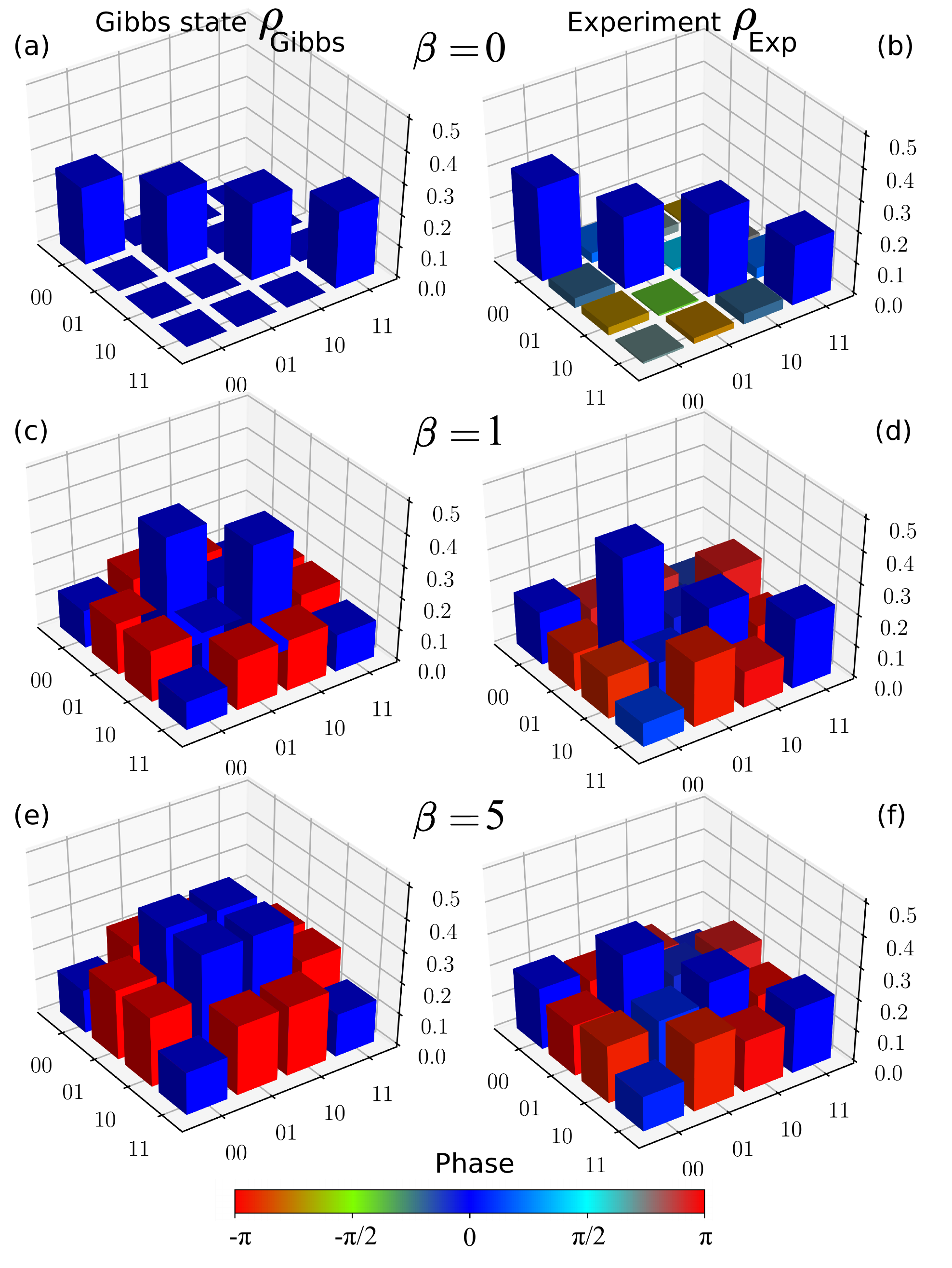}
\end{center}
\caption{
\label{fig:Label_FIG4}
\textbf{Qualitative comparison of optimized states to the Gibbs state.}
Visualization of the density matrices (in the computational basis) for the targeted Gibbs states $\rhoGibbs$ (left) and the optimized experimental states $\rhoexp$ (right) at (a-b) $\invTemp=0$, (c-d) $\invTemp=1$ and (e-f) $\invTemp=5$.
As $\invTemp$ increases, the Gibbs state monotonically develops coherence between all states, with phase $0$ $(\pi)$ for states with the same (opposite) parity. Populations in even (odd) parity states decrease (increase).
The optimized experimental states show qualitatively similar trends.
}
\end{figure}

The challenge faced by the variational algorithm is to balance the mixture of the states at each $\invTemp$, in order to generate the corresponding Gibbs state.
When working with small systems, it is possible and tempting to predetermine the variational parameters at each $\invTemp$ by a prior classical simulation and optimization for an ideal or noisy quantum processor.  We refer to this common practice~\cite{Zhu19,Francis20} as \emph{cheating}, since this approach does not scale to larger problem sizes and skips the main quality of variational algorithms: to arrive at the parameters variationally. Here, we avoid cheating altogether by starting at $\invTemp=0$, with initial guess the obvious optimal variational parameters for an ideal processor $(\gammavec=\alphavec=0)$, and using the experimentally optimized $\left( \alphavec,\gammavec \right)$ at the last $\invTemp$ as an initial guess when stepping $\invTemp$ in the range $\left[0, 5 \right]$ (in units of $1/g$). This approach only relies on the assumption that solutions (and their corresponding optimal variational angles) vary smoothly with $\invTemp$. At each $\invTemp$, we use the Gradient-Based Random-Tree optimizer of the \textit{scikit-optimize}~\cite{scikit-opt} Python package to minimize $\cost$, using 4096 averages per tomographic pre-rotation necessary for the calculation of $\cost$.  After $200$ iterations, the optimization is stopped.  The best point is remeasured two times, each with $16384$ averages per tomographic pre-rotation needed to perform two-qubit quantum state tomography of each system. A new optimization is then started for the next $\invTemp$, using the previous solution as the initial guess.

To begin comparing the optimized states $\rhoexp$ produced in experiment to the target Gibbs states $\rhoGibbs$, we first visualize their density matrices (in the computational basis) for a sampling of the $\invTemp$ range covered (\cref{fig:Label_FIG4}).
Starting from the maximally-mixed state $\II/4$ at $\invTemp=0$, the Gibbs state monotonically develops coherences (off-diagonal terms) between all states as $\invTemp$ increases. Coherences between states of equal (opposite) parity have $0$ $(\pi)$ phase throughout. Populations (diagonal terms) monotonically decrease (increase) for even (odd) parity states. By $\invTemp=5$, the Gibbs state is very close to the pure state $\ket{\groundState}\bra{\groundState}$, where $\ket{\groundState}\approx \sqrt{0.36}\left( \ket{01} + \ket{10} \right)-\sqrt{0.14}\left( \ket{00} + \ket{11} \right)$.
The noted trends are reproduced in $\rhoexp$. However, the matching is evidently not perfect, and to address this we proceed to a quantitative analysis.

\begin{figure}
\begin{center}
\includegraphics[trim={0 0.7cm 0 0},clip,width=0.5\textwidth]{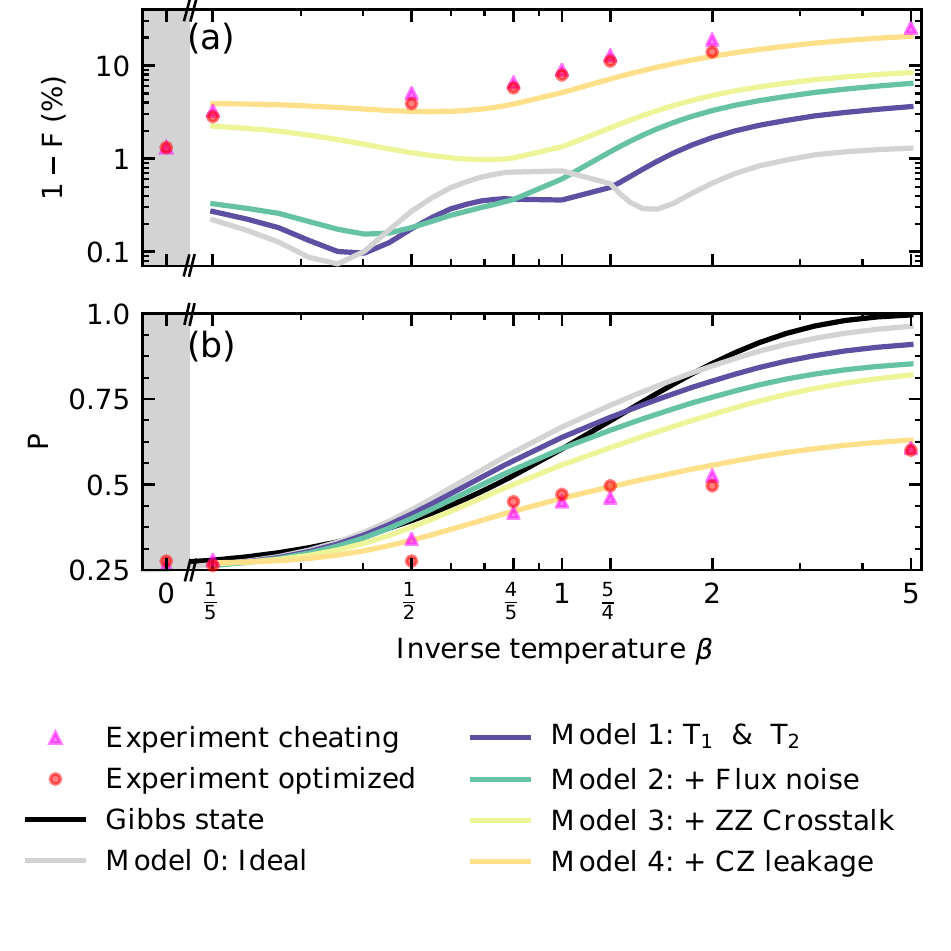}
\end{center}
\caption{
\label{fig:Label_FIG5}
\textbf{Performance of the variational algorithm.}
(a) Fidelity to the Gibbs state as a function of inverse temperature $\invTemp$ for experimental states obtained by optimization and cheating.
(b) Purity of experimental states as a function of $\invTemp$, and comparison to the purity of the Gibbs state. Added curves in both panels are obtained by numerical simulation of a noisy quantum processor with incremental error models based
on calibrated error sources for our device: qubit relaxation and dephasing times, increased dephasing from flux noise during CZ gates, residual $ZZ$ crosstalk at the bias point, and leakage during CZ gates. Leakage is identified as the dominant error source.
}
\end{figure}

We employ two metrics to quantify experimental performance: the fidelity $F$ of $\rhoexp$ to $\rhoGibbs$ and the purity $P$ of $\rhoexp$, given by
\begin{align}
\begin{split}
F&=\tr\!\left(\sqrt{ \rhoGibbs^{1/2} \rhoexp^{\vphantom{1/2}} \rhoGibbs^{1/2} }\right),\\
P&=\tr\! \left( \rhoexp^{2} \right).
\end{split}
\end{align}
At $\invTemp=0$, $F=99\%$ and $P=0.262$, revealing a very close match to the ideal maximally-mixed state. However, $F$ smoothly worsens with increasing $\invTemp$, decreasing to $92\%$ at $\invTemp=1$ and $75\%$ by $\invTemp=5$. Simultaneously, $P$ does not closely track the increase of purity of the Gibbs state. By $\invTemp=5$, the Gibbs state is nearly pure, but $P$ peaks at $0.601$.

In an effort to quantitatively explain these discrepancies, we perform a full density-matrix simulation of a four-qutrit system using \textit{quantumsim}~\cite{Quantumsim}. Our simulation incrementally adds calibrated errors for our NISQ  processor, starting from an ideal processor (model 0): transmon relaxation and dephasing times at the bias point (model 1), increased dephasing from flux noise during CZ gates (model 2), crosstalk from residual $ZZ$ coupling at the bias point (model 3), and transmon leakage to the second-excited state during CZ gates (model 4). The experimental input parameters for each increment are detailed in the Methods section and \cite{SOM_QAOA}. The added curves in \cref{fig:Label_FIG5} clearly show that model 4 quantitatively matches the observed dependence of $F$ and $P$ over the full $\invTemp$ range, and identifies leakage from CZ gates as the dominant error.

\section*{Discussion}
The power of variational algorithms relies on their adaptability: the optimizer is meant to find its way through the variational parameter space, adapting to mitigate coherent errors as allowed by the chosen parametrization. For completeness, we compare in \cref{fig:Label_FIG5} the performance achieved with our variational strategy to that achieved by cheating, i.e., using the pre-calculated optimal $\left(\alphavec, \gammavec\right)$ for an ideal processor. Our variational approach, whose sole input is the obvious initial guess at $\invTemp=0$, achieves comparable performance at all $\invTemp$. This aspect is crucial when considering the scaling with problem size, as classical pre-simulations will require prohibitive resources beyond $\sim50$ qubits, but variational optimizers would not. Given the dominant role of leakage as the error source, which cannot be compensated by the chosen  parametrization, it is unsurprising in hindsight that both approaches yield nearly identical performance.

In summary, we have presented the first generation of finite-temperature Gibbs states in a quantum computer by variational targeting of TFD states in a hybrid quantum-classical manner.
The algorithm successfully prepares mixed states for the transverse field Ising model with Gibbs-state fidelity ranging from $99\%$ to $75\%$ as $\invTemp$ increases from 0 to $5/g$. The loss of fidelity with decreasing simulated temperature is quantitatively matched by a numerical simulation with incremental error models based on experimental input parameters, which identifies leakage in CZ gates as dominant. This work demonstrates the suitability of variational algorithms on NISQ processors for the study of finite-temperature problems of interest, ranging from condensed-matter physics to cosmology. Our results also highlight the critical importance of continuing to reduce leakage in two-qubit operations when employing weakly-anharmonic multi-level systems such as the transmon.

During the preparation of this manuscript, we became aware of related experimental work~\cite{Francis20} on a trapped-ion system, applying a non-variationally prepared TFD state to the calculation of a critical point.

\section*{Methods}
We map the theoretical circuit in \cref{fig:Label_FIG1}(b) to an equivalent circuit conforming to the native gate set in our control architecture and exploiting virtual $Z$-gate compilation~\cite{McKay17} to minimize circuit depth.  Single-qubit rotations $\Rmw(\phi,\theta)$, by arbitrary angle $\theta$ around any equatorial axis $\cos(\phi)\hat{x}+\sin(\phi)\hat{y}$ on the Bloch sphere, are realized using $20~\ns$ DRAG pulses~\cite{Motzoi09, Chow10b}. Two-qubit CZ gates are realized by baseband flux pulsing~\cite{Strauch03, DiCarlo09} using the Net Zero scheme~\cite{Rol19, rol20}, completing in $80~\ns$. In the optimized circuit [\cref{fig:Label_FIG2}(b)], CZ gates only appear in pairs. These pairs are simultaneously executed and tuned as one block. Single-qubit rotations $R_1$-$R_4$ are used to change the measurement bases, as required to measure $\cost$ during optimization and to perform two-qubit tomography~\cite{Sagastizabal19} in each system to extract $F$ and $P$. A summary of single- and two-qubit gate performance and a step-by-step derivation of the optimized circuit are provided in~\cite{SOM_QAOA}.

The models used to simulate the performance of the algorithm are incremental: model $k$ contains all the noise mechanisms in model $k-1$ plus one more, which we use for labeling in \cref{fig:Label_FIG5}.
Model $0$ corresponds to an ideal quantum processor without any error. Model $1$ adds the measured relaxation and dephasing times measured for the four transmons at their bias point. These times are tabulated in~\cite{SOM_QAOA}.
Model $2$ adds the increased dephasing that flux-pulsed transmons experience during CZ gates. For this we extrapolate the echo coherence time
$\Ttwoecho$ to the CZ flux-pulse amplitude using a $1/f$ noise model~\cite{Wellstood1987,Paladino14} with amplitude $\sqrt{A} = 1 \mu\Phi_0$. This noise model is implemented following~\cite{Varbanov20}.
Model $3$ adds the idling crosstalk due to residual $ZZ$ coupling between transmons. This model expands on the implementation of idling evolution used for coherence times: the circuit gates are simulated to be instantaneous, and the idling evolution of the system is trotterized. In this case, the residual $ZZ$ coupling operator uses the measured residual $ZZ$ coupling strengths at the bias point~\cite{SOM_QAOA}. Finally, model $4$ adds leakage to the CZ gates, based on randomized benchmarking with modifications to quantify leakage \cite{Asaad16,Rol19}, and implemented in simulation using the procedure described in~\cite{Varbanov20}.

Leakage to transmon second-excited states is found essential to quantitatively match the performance of the algorithm by simulation.
To reach this conclusion it was necessary to first thoroughly understand how leakage affects the two-qubit tomographic reconstruction procedure employed.
The readout calibration only considers computational states of the two transmons involved. Moreover, basis pre-rotations only act on the qubit subspace, leaving the population in leaked states unchanged.
Using an overcomplete set of basis pre-rotations for state tomography, comprising both positive $\left( X,Y, Z\right)$ and negative $\left(-X,-Y,-Z\right)$ bases for each transmon, leads to the misdiagnosis of a leaked state as a maximally mixed state qubit state for that transmon. This is explained in~\cite{SOM_QAOA}.

%%%% ACKNOWLEDGMENTS %%%%%
\begin{acknowledgments}
We thank L.~Janssen, M.~Sarsby, and M.~Venkatesh for experimental assistance, X.~Bonet-Monroig and B.~Tarasinski for useful discussions, and G.~Calusine and W.~Oliver for providing the travelling-wave parametric amplifier used in the readout amplification chain. This research is supported by Intel Corporation, the ERC Synergy Grant QC-lab, and by the Office of the Director of National Intelligence (ODNI), Intelligence Advanced Research Projects Activity (IARPA), via the U.S. Army Research Office grant W911NF-16-1-0071. The views and conclusions contained herein are those of the authors and should not be interpreted as necessarily representing the official policies or endorsements, either expressed or implied, of the ODNI, IARPA, or the U.S. Government.
\end{acknowledgments} 

%%%%%%%%% REFERENCES 1 %%%%%%%%%

\end{bibunit}
\begin{bibunit}[apsrev4-2]
\clearpage
\renewcommand{\theequation}{S\arabic{equation}}
\renewcommand{\thefigure}{S\arabic{figure}}
\renewcommand{\thetable}{S\arabic{table}}
\renewcommand{\bibnumfmt}[1]{[S#1]}
\renewcommand{\citenumfont}[1]{S#1}
\setcounter{section}{0}
\setcounter{figure}{0}
\setcounter{equation}{0}
\setcounter{table}{0}

\parskip \baselineskip
\def\myupbracefill#1{\rotatebox{90}{\stretchto{\{}{#1}}}
\def\rlwd{.5pt}

\newcommand\notate[4][B]{%
  \if B#1\else\def\myupbracefill##1{}\fi%
  \def\useanchorwidth{T}%
  \setbox0=\hbox{$\displaystyle#2$}%
  \def\stackalignment{c}\stackunder[-6pt]{%
    \def\stackalignment{c}\stackunder[-1.5pt]{%
      \stackunder[2pt]{\strut $\displaystyle#2$}{\myupbracefill{\wd0}}}{%
    \rule{\rlwd}{#3\baselineskip}}}{%
  \strut\kern9pt$\rightarrow$\smash{\rlap{$~\displaystyle#4$}}}%
}

\onecolumngrid
\section*{Supplemental material for 'Variational preparation of finite-temperature states on a quantum computer'}
\date{\today}
\maketitle

This supplement provides additional information in support of statements and claims made in the main text.
Section~\ref{sec:one} presents the optimization of the engineered cost function.
Section~\ref{sec:two} provides a step-by-step description of the transformation of the circuit in \cref{fig:Label_FIG1}(b) into the equivalent, optimized circuit in \cref{fig:Label_FIG2}(b).
Section~\ref{sec:three} provides further information on the device and transmon parameters measured at the bias point.
Section~\ref{sec:four} presents a detailed description of the fridge wiring and electronic-control setup.
Section~\ref{sec:five} summarizes single- and two-qubit gate performance.
Section~\ref{sec:six} characterizes residual $ZZ$ coupling at the bias point.
Section~\ref{sec:seven} details the measurement procedures used for cost-function evaluation and for two-qubit state tomography.
Section~\ref{sec:eight} explains the impact of transmon leakage on two-qubit tomography.
Section~\ref{sec:nine} explains the package and error models used in the numerical simulation.

\section{Optimization of the cost function}
\label{sec:one}
\begin{figure}
\begin{center}
   \includegraphics[width=\columnwidth]{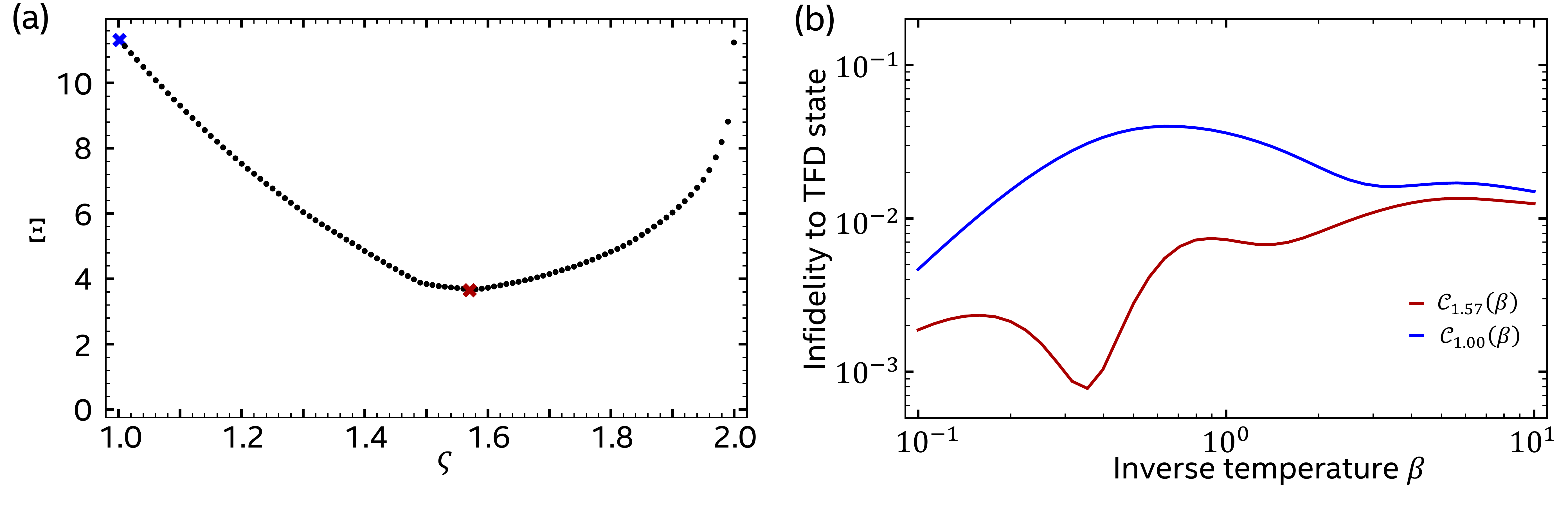}
 \end{center}
\caption{
\label{fig:cost_function_variation}
Optimization of the cost function.
(a) Plot of $\Xi$ versus $\varsigma$. The minimum of $\Xi$ is found near $\varsigma = 1.57$, indicating the closest match between optimized states and ideal TFD states across the range of $\invTemp$ considered. (b) Simulated infidelity to the ideal TFD state of states optimized using the cost function $\cost_{1.00}$ (blue) and the optimized cost function $\cost_{1.57}$ (red).
}
\end{figure}

We optimize the cost function to maximize fidelity of the variationally-optimized state $\ket{\psi(\alphavec,\gammavec)}$ to the TFD state $\TFD$ (assuming an ideal processor).
We consider the class of cost functions defined by
\begin{align}
\cost_\varsigma(\invTemp) &= \XA + \XB + \varsigma (\ZZA + \ZZB) -\invTemp^{-\varsigma} (\XXAB + \ZZAB),
\label{eq:cost_function_varsigma}
\end{align}
and perform nested optimization of parameter $\varsigma$ to minimize the infidelity of the variationally-optimized state to the TFD state over a range of inverse temperatures
\begin{align}
\mathcal{B} = \left \{ 10^{x/2} : \left(x \in \mathbb{Z}\right) \wedge  \left( -8 \leq x \leq 8\right) \right \}.
\end{align}
We define the minimization quantity of interest as
\begin{align}
\Xi(\varsigma) &= \sum_{\invTemp \in \mathcal{B}} \sum_{o \in \mathcal{O}} \left\vert \expect{\mathrm{TFD}(\invTemp)}{o}{\mathrm{TFD}(\invTemp)} - \expect{\optimizedState(\varsigma,\invTemp)}{o}{\optimizedState(\varsigma,\invTemp)} \right\vert,
\end{align}
where $\mathcal{O}$ is the set of operators
\[
\mathcal{O} = \left\{ \XA, \YA, \ZA, \XB, \YB, \ZB, \XXA, \YYA, \ZZA, \XXB, \YYB, \ZZB, \XXAB, \YYAB, \ZZAB\right\},
\]
and $\ket{\optimizedState(\varsigma,\invTemp)}$ is the state optimized using $\cost_\varsigma(\invTemp)$.
We find the minimum value of $\Xi$ at $\varsigma=1.57$ [see \cref{fig:cost_function_variation}(a)].

\begin{figure}
\begin{center}
   \includegraphics[width=\columnwidth]{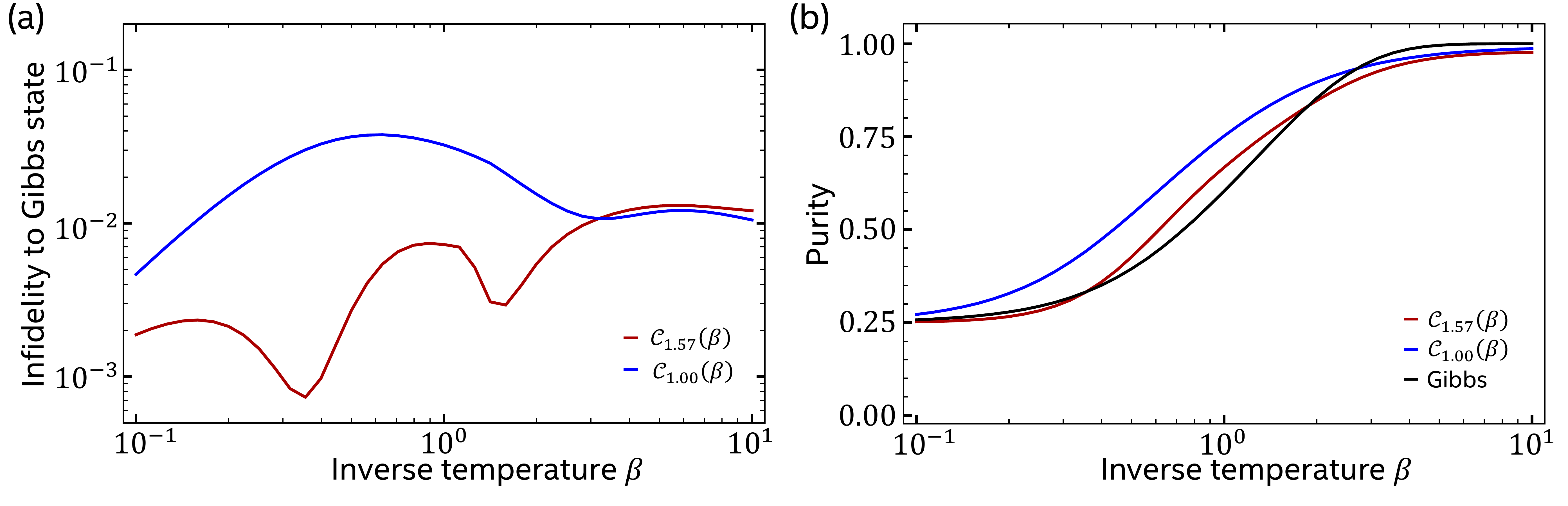}
 \end{center}
\caption{
\label{fig:cost_function_comparison}
Cost function performance comparison.
(a) Simulated infidelity of the reduced state of system A to the Gibbs state for states optimized using the cost function $\cost_{1.00}$ (blue)  and the optimized cost function $\cost_{1.57}$ (red).
(b) Corresponding purity of the reduced state. The purity of the Gibbs state is also shown for comparison (black).
}
\end{figure}

We compare the performance of the optimized cost function $\cost_{1.57}$ to that used in prior work, $\cost_{1.00}$, in two ways. First, we compare the simulated infidelity to the TFD state of states optimized with both cost functions in the range $\invTemp\in[0.1,10]$.
The optimized cost function $\cost_{1.57}$ performs better over the entire range. Finally, we compare the simulated fidelity $F$ of the reduced state of system A to the targeted Gibbs state. As shown in \cref{fig:cost_function_comparison}, using $\cost_{1.57}$ significantly reduces the infidelity $1-F$ for $\invTemp\lesssim3$, We also observe that the purity of the reduced state tracks that of the Gibbs state more closely when using $\cost_{1.57}$.

\section{Circuit compilation}
\label{sec:two}
In this section we present the step-by-step transformation of the circuit in \cref{fig:Label_FIG1}(b) into the equivalent circuit in \cref{fig:Label_FIG2}(b) realizable with the native gate set in our control architecture.
\vspace{0.2cm}

\begin{figure*}[h!]
\begin{center}
   \includegraphics[width=0.9\textwidth]{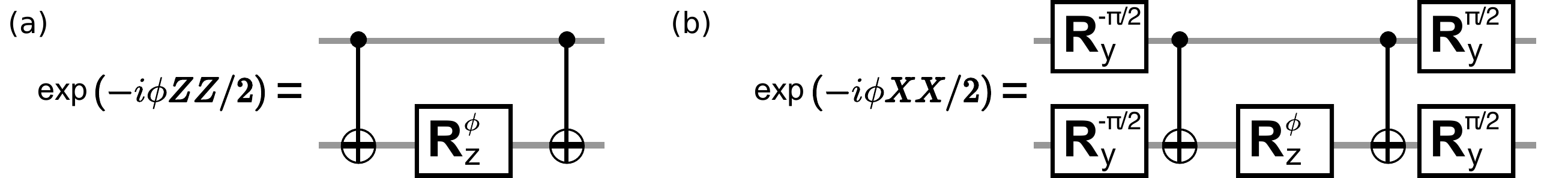}
 \end{center}
\caption{
\label{fig:ZZandXX}
Exponentiation of $\ZZ$ and $\XX$.
(a) Standard compilation of $e^{-i\phi \ZZ/2}$ using CNOT gates and single-qubit rotations. (b) Standard compilation of $e^{-i \phi \XX/2}$ using additional basis pre- and post-rotations.
}
\label{fig:compilation_step_1}
\end{figure*}

\noindent\textbf{Exponentiation of $\ZZ$ and $\XX$:} We first substitute the standard decomposition of the operations $e^{-i\phi \ZZ/2}$ and $e^{-i\phi \XX/2}$  using controlled-NOT (CNOT) gates and single-qubit rotations, shown in \cref{fig:ZZandXX}.
The decomposition of $e^{-i\phi \ZZ/2}$ uses an initial CNOT to transfer the two-qubit parity into the target qubit, followed by a rotation $\Rz(\phi)$ on this target qubit, and a final CNOT inverting the parity.
The decomposition of $e^{-i\phi \XX/2}$ simply dresses the transformations above by pre- and post-rotations transforming from the $X$ basis to the $Z$ basis and back, respectively. The result of these substitutions is shown in \cref{fig:compilation_step_1}.

\begin{figure*}[h!]
\begin{center}
   \includegraphics[width=\textwidth]{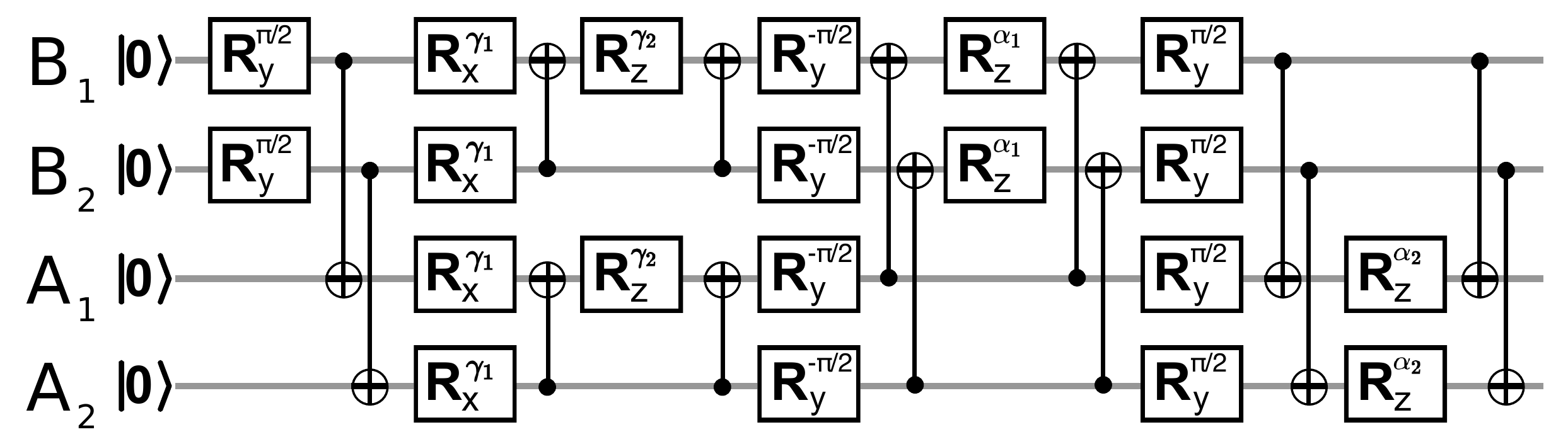}
 \end{center}
\caption{
\label{fig:compilation_step_1}
Compilation step 1.
Depth-14 circuit obtained by replacing the $\ZZ$ and $\XX$ exponentiation steps in \cref{fig:Label_FIG1}(b) with the circuits of \cref{fig:ZZandXX}.
}
\end{figure*}

\begin{figure*}[h!]
\begin{center}
   \includegraphics[width=0.4\textwidth]{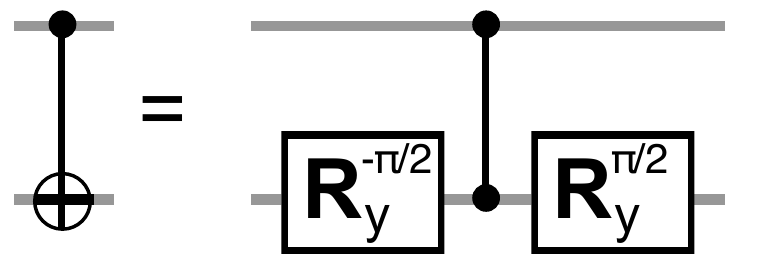}
 \end{center}
\caption{Compilation of CNOT gate using native gates: one CZ sandwiched by single-qubit rotations on the target qubit.}
\label{fig:cnot_to_cphase}
\end{figure*}

\begin{figure*}[h!]
\begin{center}
   \includegraphics[width=\textwidth]{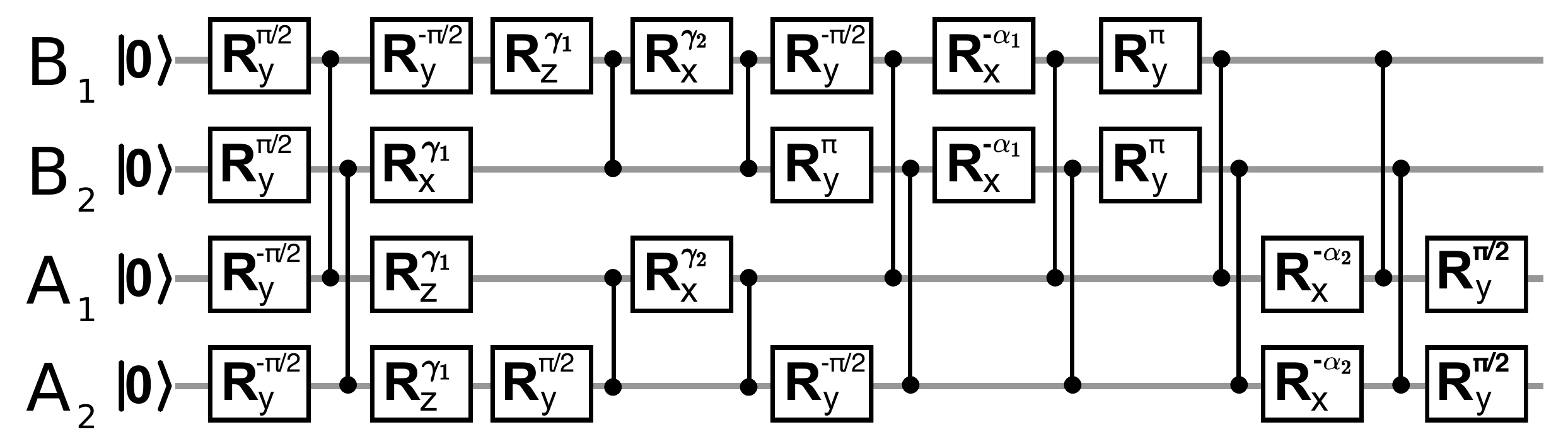}
 \end{center}
\caption{
\label{fig:compilation_step_2}
Compilation step 2.
Depth-16 circuit obtained by replacing every CNOT in \cref{fig:compilation_step_1} with the circuit of \cref{fig:cnot_to_cphase} and using simple identities.
}
\end{figure*}

\vspace{0.2cm}
\noindent\textbf{Compilation using native gate set:} The native gate set consists of single-qubit rotations of the form $\Rmw(\phi,\theta)$ and CZ gates. We compile every CNOT in \cref{fig:compilation_step_1} as a circuit using native gates, shown in \cref{fig:cnot_to_cphase}. Note that
$\Ry(\theta)=\Rmw(90\degrees, \theta)$. Using this replacement together with the identities $\Ry(-90\degrees)\Rz(\phi)\Ry(90\degrees)=\Rx(-\phi)$ and $\Ry(-90\degrees)\Rx(\phi)\Ry(90\degrees)=\Rz(\phi)$ leads to the circuit in \cref{fig:compilation_step_2}.

\begin{figure*}[h!]
\begin{center}
   \includegraphics[width=0.4\textwidth]{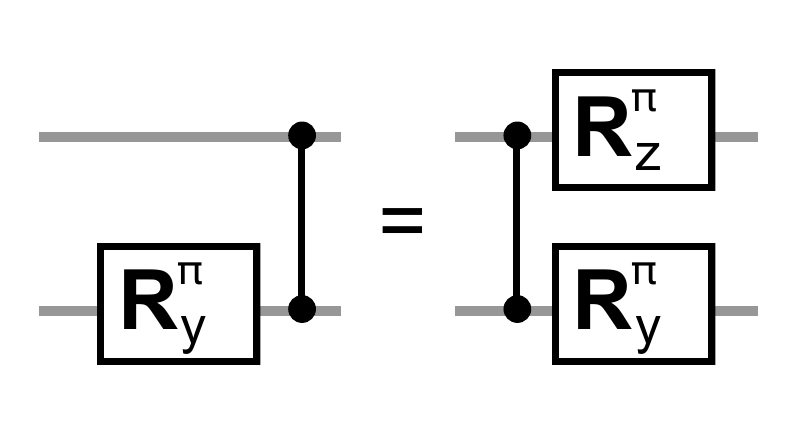}
 \end{center}
\caption{
\label{fig:CPhase_Ygate}
Commutation of $\Ry(180\degrees)$ and CZ gates.
}
\end{figure*}

\begin{figure*}[h!]
\begin{center}
   \includegraphics[width=0.8\textwidth]{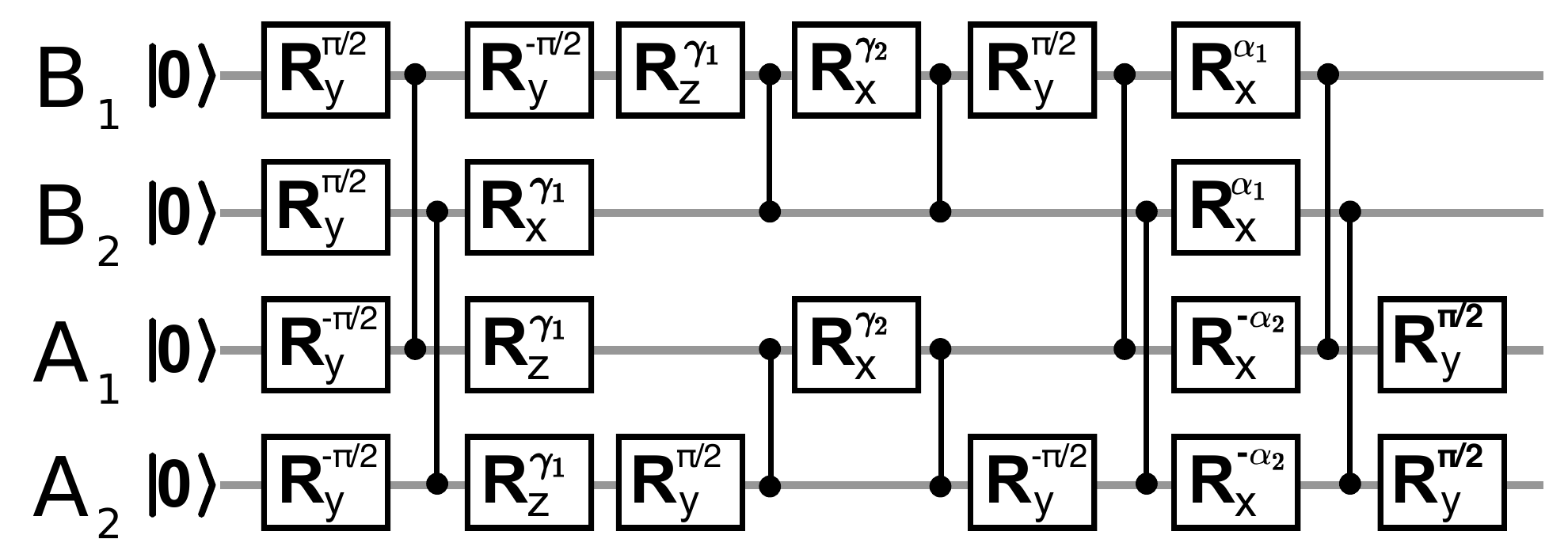}
 \end{center}
\caption{
\label{fig:compilation_step_3}
Compilation step 3.
Depth-12 circuit obtained by applying the commutation rule in \cref{fig:CPhase_Ygate} and simple identities to the circuit of \cref{fig:compilation_step_2}.
}
\end{figure*}

\vspace{0.2cm}
\noindent\textbf{Reduction of circuit depth:} Exploiting the commutations in \cref{fig:CPhase_Ygate} together with the identities $\Ry(-90\degrees) = \Ry(180\degrees)\Ry(90\degrees)$ and $\Ry(180\degrees)\Rx(\phi)\Ry(180\degrees)=\Rx(-\phi)$, we can bring two identical pairs of CZ gates back-to-back
and cancel them out (since CZ is its own inverse). This leads to the circuit in \cref{fig:compilation_step_3}.

\begin{figure*}[h!]
\begin{center}
   \includegraphics[width=0.8\textwidth]{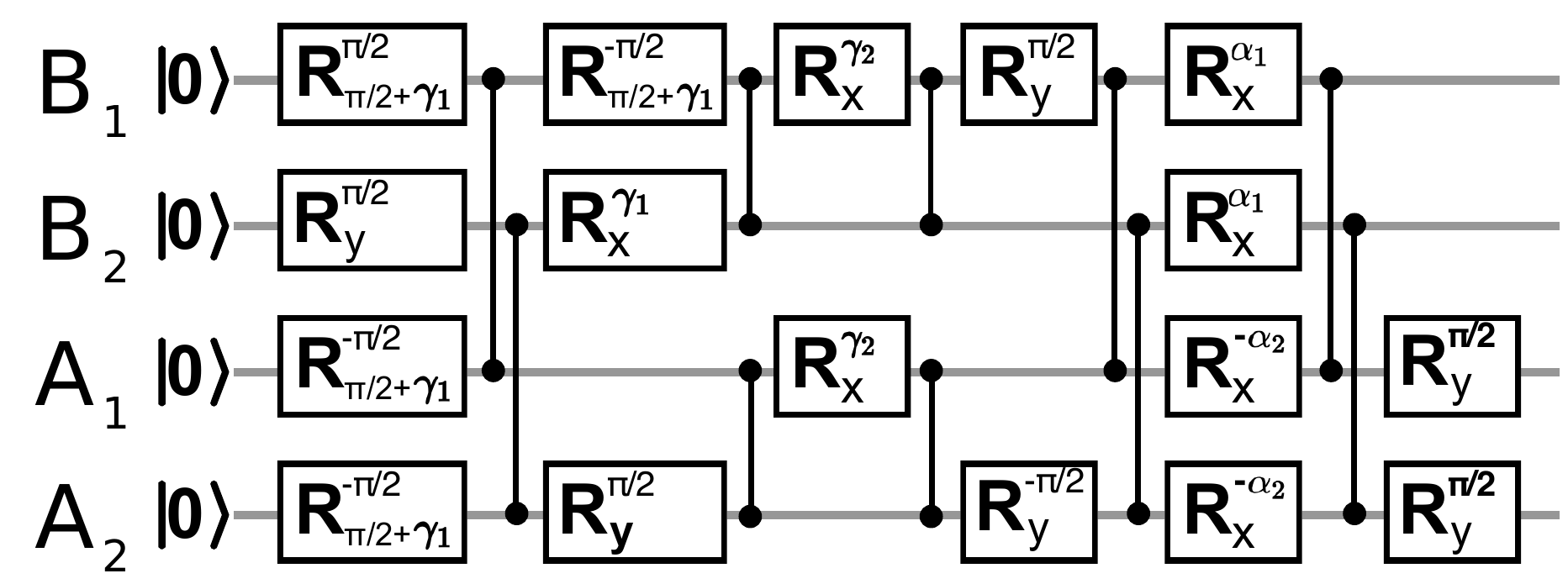}
 \end{center}
\caption{
\label{fig:compilation_step_4}
Compilation step 4.
Depth-11 circuit obtained by propagating all $\Rz$ gates in \cref{fig:compilation_step_3} to the beginning of the circuit and then eliminating them.
}
\end{figure*}

\vspace{0.2cm}
\noindent\textbf{Elimination of $Z$ rotations:} All the $\Rz$ gates in \cref{fig:compilation_step_3} can be propagated to the beginning of the circuit using the commutation relation
\[
    \Rz(\alpha) \Rmw(\phi,\theta)= \Rmw(\phi+\alpha,\theta) \Rz(\alpha)
\]
and noting that $\Rz$ commutes with CZ. State $\ket{0}$ is an eigenstate of all $\Rz$ rotations, so we can  ignore all $\Rz$ gates at the start because they only produce a global phase.
This action leads to the final depth-11 circuit shown in \cref{fig:compilation_step_4}, which matches that of \cref{fig:Label_FIG2}(b) upon adding measurement pre-rotations and final measurements on all qubits.

\section{Device and transmon parameters at bias point}
\label{sec:three}
Our experiment makes use of four transmons with square connectivity within a seven-transmon processor.
Figure~S10 provides an optical image zoomed in to this transmon patch.
Each transmon has a flux-control line for two-qubit gating, a microwave-drive line for single-qubit gating, and dispersively-coupled resonator with Purcell filter for readout~\cite{Heinsoo18, Bultink20}.
The readout-resonator/Purcell-filter pair for $\QBone$ is visible at the center of this image. A vertically running common feedline connects to all Purcell filters, enabling simultaneous readout of the four transmons by frequency multiplexing.
Air-bridge crossovers enable the routing of all input and output lines to the edges of the chip, where they connect to a printed circuit board through aluminum wirebonds. The four transmons are biased to their sweetspot using static flux bias to counter any residual offset. Table~S1 presents measured transmon parameters at this bias point.

\begin{figure*}[h!]
 \begin{center}
   \includegraphics[width=0.7\textwidth]{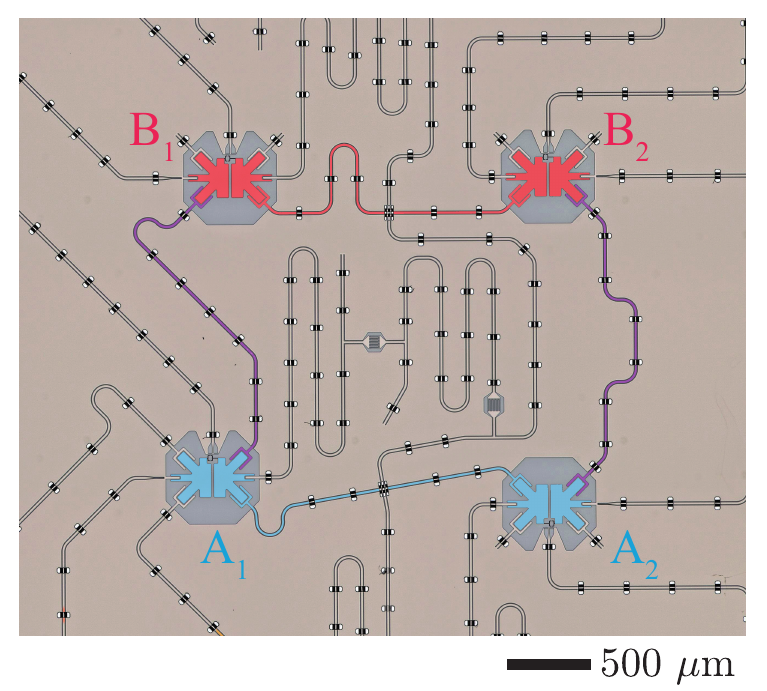}
 \end{center}
 \caption{
 Optical image of the device, zoomed in to the four transmons used in the experiment.
 Added false color highlights the transmon pair of system A (blue, $\QAone$, $\QAtwo$) the transmon pair of system B (red, $\QBone$, $\QBtwo$), and the dedicated bus resonators used to achieve intra-system  (read and blue) and inter-system coupling (purple).
 }
\end{figure*}

\begin{table}[h!]
\begin{tabular}{ l | c | c | c | c }
\hline
Transmon & $\QBone$ & $\QBtwo$ & $\QAone$ & $\QAtwo$\\
\hline
\hline
Sweetspot frequency ($\GHz$) & $6.433$ & $5.771$ & $5.887$ & $4.534$\\
Relaxation time $\Tone$ $\left(\us\right)$ & $32.1$ & $40.7$ & $64.0$ & $33.7$\\
Echo dephasing time $\Ttwoecho$ $\left(\us\right)$ & $29.9$ & $40.5$ & $45.9$ & $68.8$\\
\hline
Readout frequency ($\GHz$) & $7.493$ & $7.225$ & $7.058$ & $6.913$\\
Average assignment fidelity ($\%$) & $96.5$ & $96.5$ & $97.0$ & $93.8$\\
\hline
\end{tabular}
\caption{Summary of measured transmon parameters at bias point.}
  \label{table:specs}
\end{table}

\section{Experimental setup}
\label{sec:four}
The device was mounted on a copper sample holder attached to the mixing chamber of a Bluefors XLD dilution refrigerator with $12~\mK$ base temperature.
For radiation shielding, the cold finger was enclosed by a copper can coated with a mixture of Stycast 2850 and silicon carbide granules ($15\mathrm{-}1000~\nm$ diameter) used for infrared absorption.
To shield against external magnetic fields, the can was enclosed by an aluminum can and two Cryophy cans.
Microwave-drive lines were filtered using $\sim60~\dB$ of attenuation with both commercial cryogenic attenuators and home-made Eccosorb filters for infrared absorption.
Flux-control lines were also filtered using commercial low-pass filters and Eccosorb filters with stronger absorption.
Flux pulses for CZ gates were coupled to the flux-bias lines via room-temperature bias tees.
Amplification of the readout signal was done in three stages: a travelling-wave parametric amplifier (TWPA, provided by MIT-LL~\cite{Macklin15}) located at the mixing chamber plate, a Low Noise Factory HEMT at the $4~\K$ plate, and finally a Miteq amplifier at room temperature.

Room-temperature electronics used both commercial hardware and custom hardware developed in QuTech.
Rohde \& Schwarz SGS100 sources provided all microwave signals for single-qubit gates and readout.
Home-built current sources (IVVI racks) provided static flux biasing.
QuTech arbitrary waveform generators (QWG) generated the modulation envelopes for single-qubit gates and a Zurich Instruments HDAWG-8 generated the flux pulses for CZ gates.
A Zurich Instruments UHFQA was used to perform independent readout of the four qubits.
QuTech mixers were used for all frequency up- and down-conversion.
The QuTech Central Controller (QCC) coordinated the triggering of the QWG, HDAWG-8 and UHFQA.

All measurements were controlled at the software level with QCoDeS~\cite{QCoDeS16} and PycQED~\cite{PycQED16} packages.
The QuTech OpenQL compiler translated high-level Python code into the eQASM code~\cite{Fu19} forming the input to the QCC.

\begin{figure*}[h!]
\begin{center}
\includegraphics[width=\textwidth]{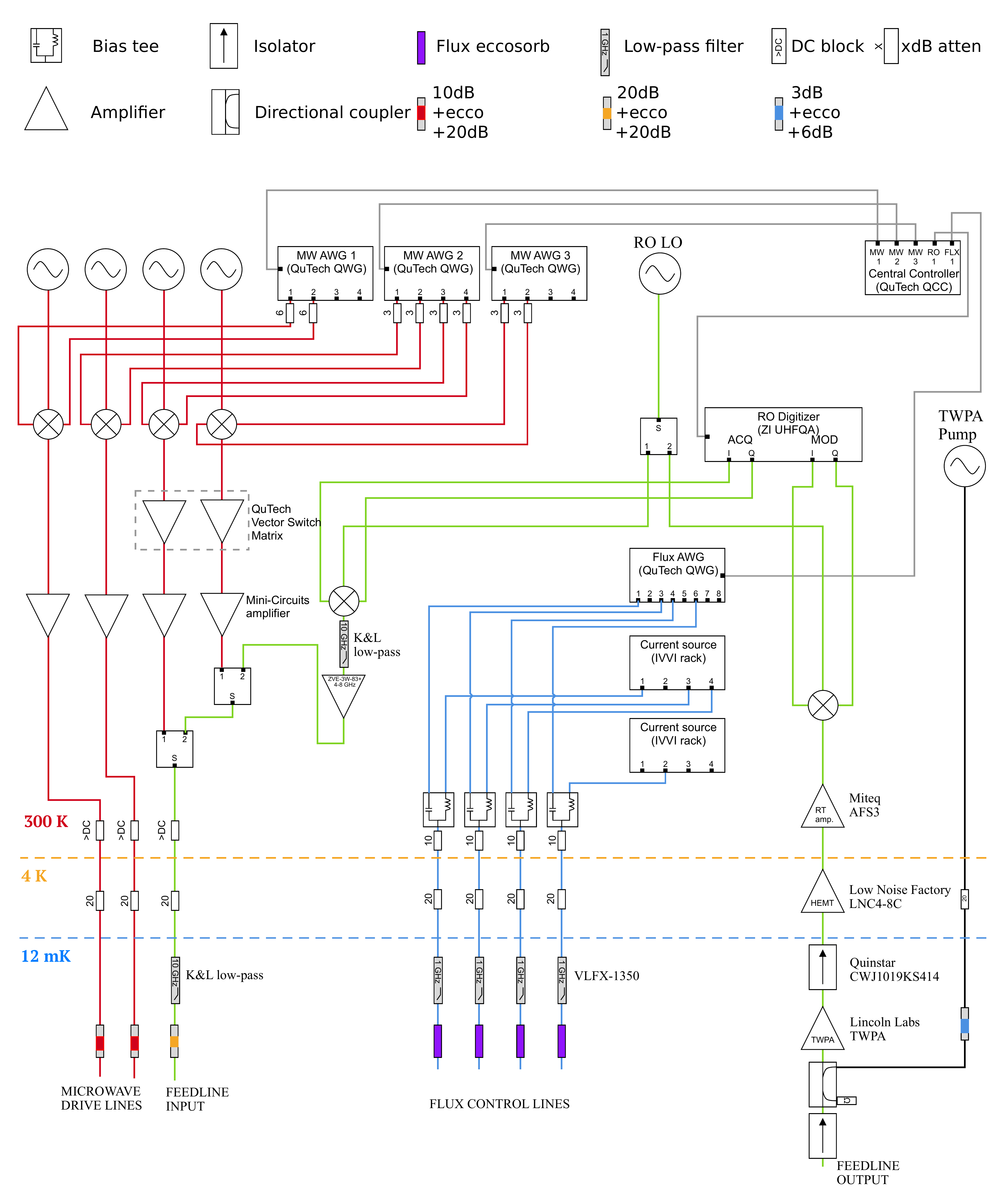}
\end{center}
\caption{
Fridge wiring and electronic setup.
Please see text for details.
}
\end{figure*}

\section{Gate performance}
\label{sec:five}

The gate set in our quantum processor consists of single-qubit rotations $\Rmw(\phi,\theta)$ and two-qubit CZ gates.
Single-qubit rotations are implemented as DRAG-type microwave pulses with total duration $4\sigma=20~\ns$, where $\sigma$ is the Gaussian width of the main-quadrature Gaussian pulse envelope.
We characterize single-qubit gate performance by single-qubit Clifford randomized benchmarking (100 seeds per run) with modifications to detect leakage, keeping all other qubits in $\ket{0}$.
Two-qubit CZ gates are implemented using the Net Zero flux-pulsing scheme, with strong pulses acquiring the conditional phase in $70~\ns$ and weak pulses nulling single-qubit phases in $10~\ns$.
Intra-system and inter-system CZ gates were simultaneously tuned in pairs (using conditional-oscillation experiments as in~\cite{Rol19}) in order to reduce circuit depth.
However, we characterize CZ gate performance individually using two-qubit interleaved randomized benchmarking (100 seeds per run) with modifications to detect leakage, keeping the other two qubits in $\ket{0}$.
Figure~S12 presents the extracted infidelity and leakage for single-qubit gates (circles) and CZ gates (squares).

\begin{figure}[h!]
\begin{center}
  \includegraphics[width=\columnwidth]{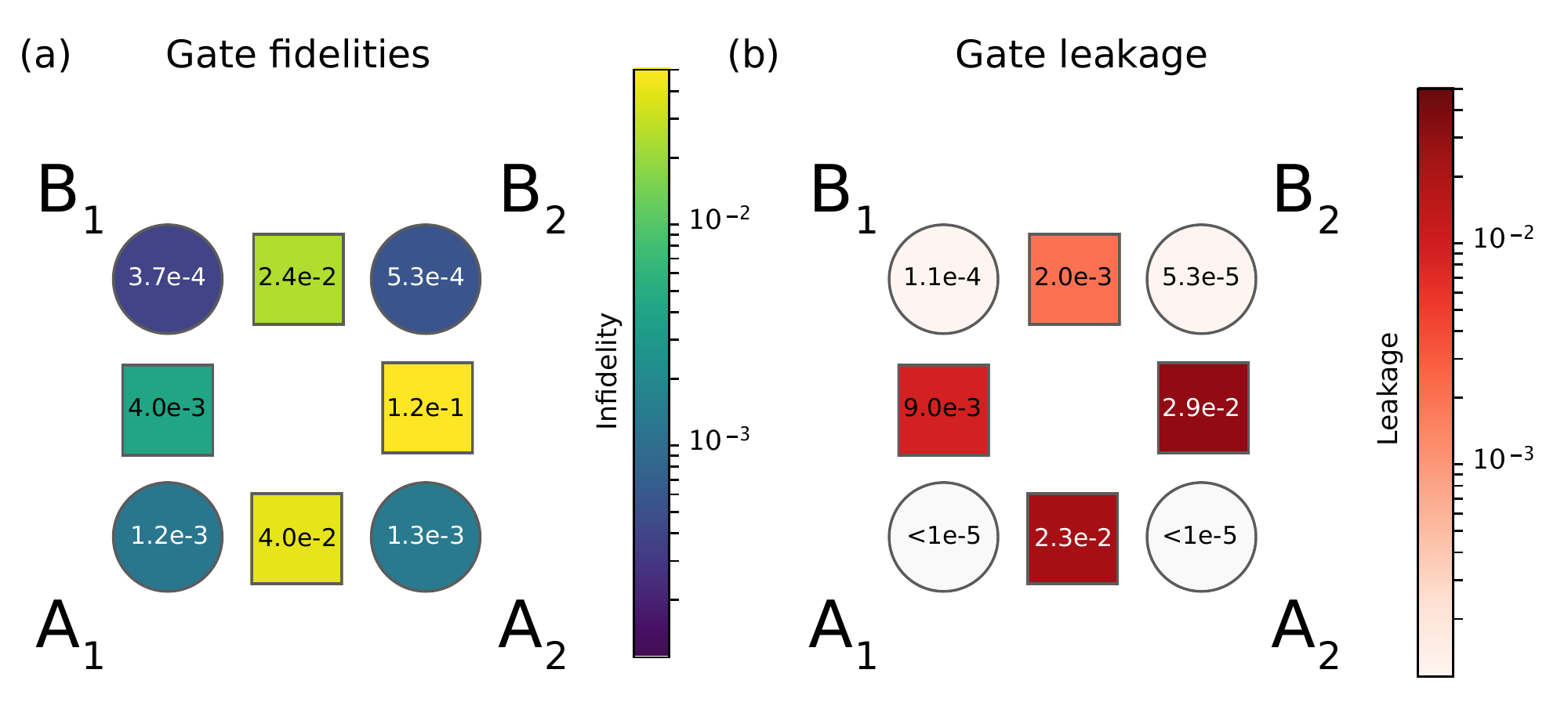}
\end{center}
\caption{
Single- and two-qubit gate performance. (a) Infidelity and (b) leakage of single-qubit gates (circles) and CZ gates (squares), extracted by randomized benchamarking.
}
\label{fig:SOM_leakagefid}
\end{figure}

\section{Residual ZZ coupling at bias point}
\label{sec:six}
\begin{figure*}
\begin{center}
\includegraphics[width=\textwidth]{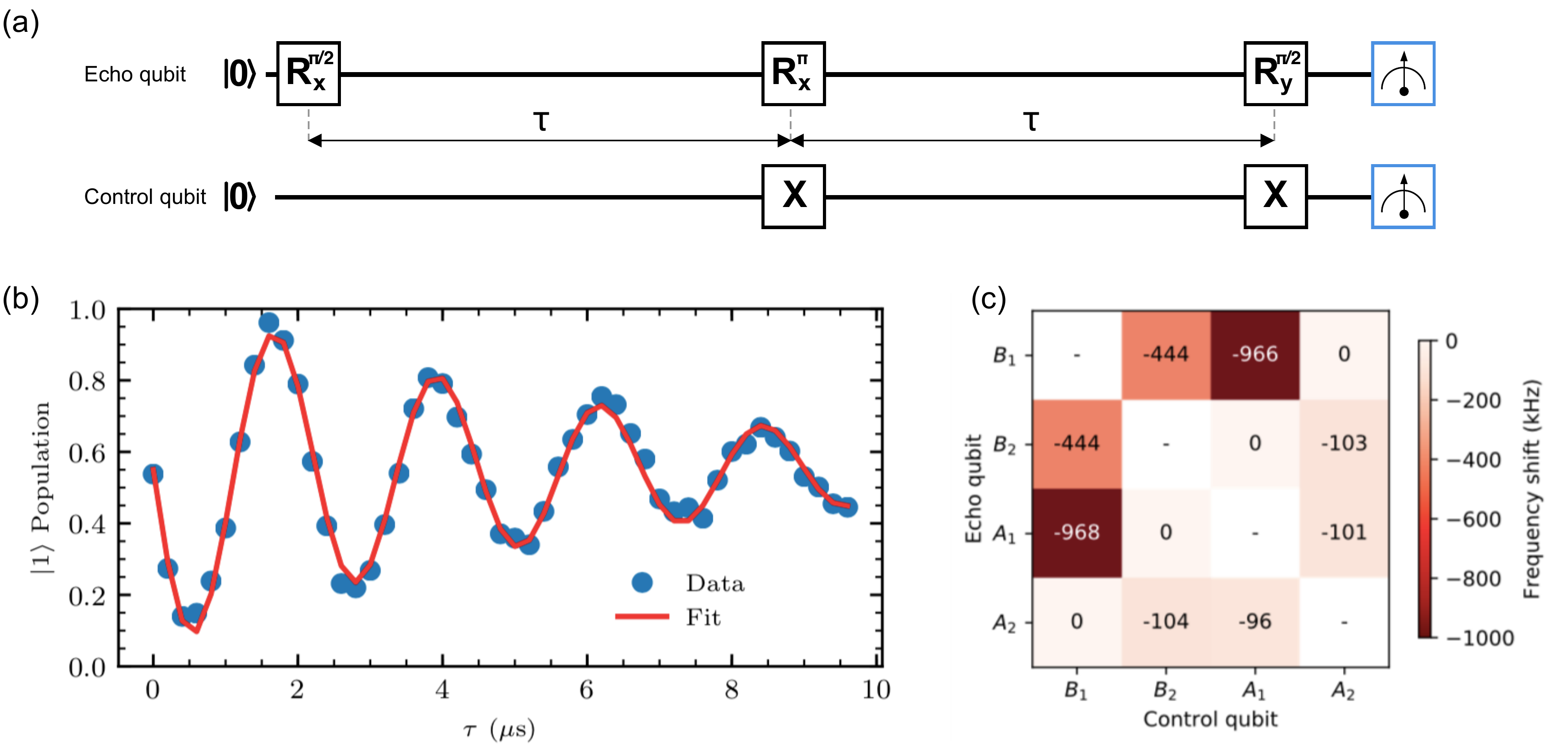}
\end{center}
\caption{
Characterization of residual $ZZ$ coupling.
(a) Modified echo experiment to determine the shift in frequency of one qubit (echo qubit) when another (the control qubit) changes state from $\ket{0}$ to $\ket{1}$.
(b) Example data for pair $\QBtwo$ (echo) - $\QBone$ (control).
(c) Table of extracted frequency shifts for all pairs of echo and control qubits.
}
\label{fig:SOM_ZZ_res}
\end{figure*}

Coupling between nearest-neighbor transmons in our device is realized using dedicated coupling bus resonators.
The non-tunability of these couplers leads to residual $ZZ$ coupling between the transmons at the bias point.
We quantify the residual $ZZ$ coupling between every pair of transmons as the shift in frequency of one when the state of the other changes from $\ket{0}$ to $\ket{1}$.
We extract this frequency shift using the simple time-domain measurement shown in \cref{fig:SOM_ZZ_res}(a): we perform a standard echo experiment on one qubit (the echo qubit), but add a $\pi$ pulse on the other qubit (control qubit) halfway through the free-evolution period simultaneous with the refocusing $\pi$ pulse on the echo qubit. An example measurement with $\QBtwo$ as the echo qubit and $\QBone$ as the control is shown in \cref{fig:SOM_ZZ_res}(b).  The complete results for all Echo-qubit, control-qubit combinations are presented as a matrix in \cref{fig:SOM_ZZ_res}(c).
We observe that the residual $ZZ$ coupling is highest between $\QBone$ and the mid-frequency qubits $\QBtwo$ and $\QAone$.  This is consistent with the higher (lower) absolute detuning and the lower (higher) transverse coupling between $\QAtwo$ ($\QBone$) and the mid-frequency transmons.

\section{Measurement models, cost function evaluation, and two-qubit state tomography}
\label{sec:seven}
\begin{figure*}[h!]
\begin{center}
\includegraphics[width=0.4\textwidth]{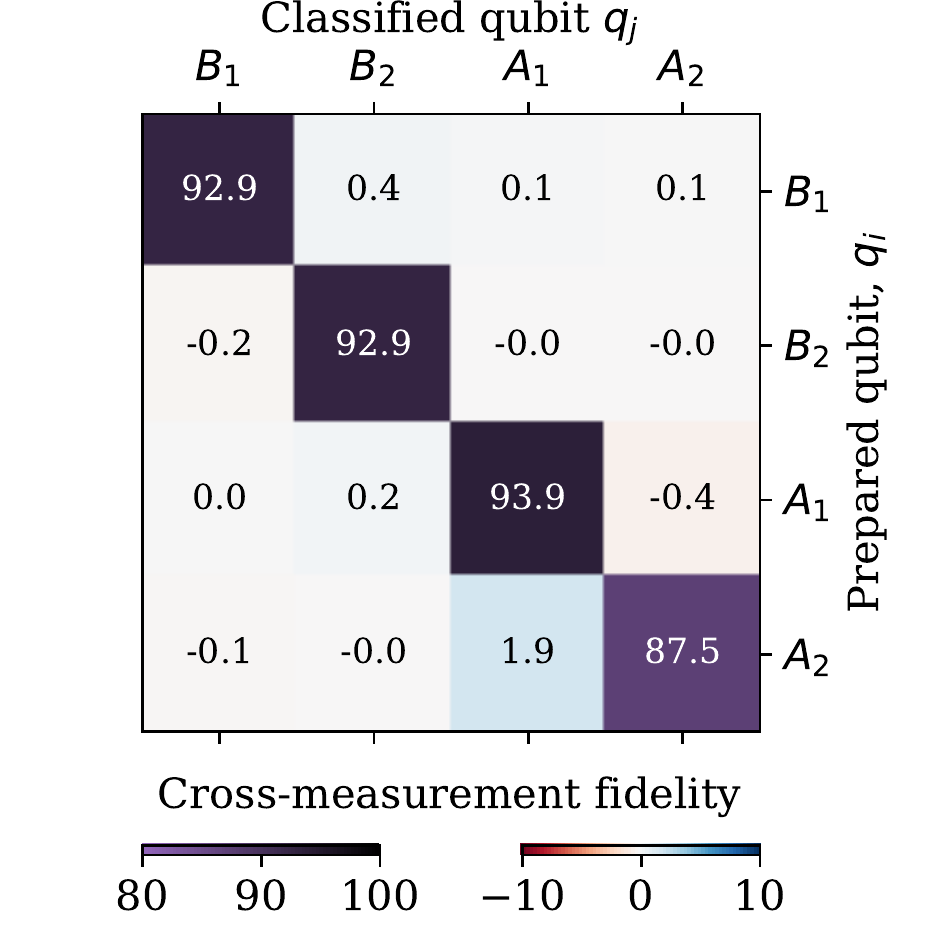}
\end{center}
\caption{Measured readout cross-fidelity matrix.}
\label{fig:SOM_crossfid}
\end{figure*}

In this section we present detailed aspects of measurement as needed for evaluation of $\cost$ and for performing two-qubit state tomography.
We begin by characterizing the fidelity and crosstalk of simultaneous single-qubit measurements using the cross-fideltiy matrix as defined in~\cite{Andersen20}:
\[
F_{ji} = 1- \mathrm{Prob}\left( e_j | I_i \right) - \mathrm{Prob}\left( g_j | \pi_i \right),
\]
where $e_j$ ($g_j$) denotes the assignment of qubit $j$ to the $\ket{1}$ $(\ket{0})$ state, and $\pi_i$ ($I_i$) denotes the preparation of qubit $i$ in $\ket{1}$ $(\ket{0})$.
The measured cross-fidelity matrix for the four qubits is shown in \cref{fig:SOM_crossfid}.
From diagonal element $F_{ii}$ we extract the average assignment fidelity for qubit $i$, the latter given by $1/2+F_{ii}/2$ and quoted in Table~S1.
The magnitude of the off-diagonal elements $F_{ji}$ with $j\neq i$ quantifies readout crosstalk, and is below $2\%$ for all pairs. This low level of crosstalk justifies using the simple measurement models that we
now describe.

\subsection{Measurement models}
The evaluation of the cost function and two-qubit state tomography require estimating the expected value of single-qubit and two-qubit Pauli operators. We do so by least-squares linear inversion of the experimental average of
single-transmon measurements and two-transmon correlation measurements.

When measuring transmon $i$, we 1-bit discretize the integrated analog signal for its readout channel at every shot, outputting  $m_i=+1$ when declaring the transmon in $\ket{0}$ and $m_i=-1$ when declared it in $\ket{1}$.
The expected value of $\langle m_i \rangle$ is given by $\overline{m}_i=\textrm{Tr}\left(M_i \rhoexp\right)$, where the measurement operator (in view of the low crosstalk) is modelled as
\[
M_i = \sum_{k=0} c_{k}^{i}\ket{k_i}\bra{k_i},
\]
where $k$ denotes transmon excitation level, and $c_k^i \in [-1,1]$ are real-valued coefficients.
Making use of the 1-qubit Pauli operators $I_i=\ket{0_i}\bra{0_i}+\ket{1_i}\bra{1_i}$ and $Z_i=\ket{0_i}\bra{0_i}-\ket{1_i}\bra{1_i}$ and truncating to three transmon levels, we can rewrite the measurement operator in the form
\begin{equation}
M_i = c_I^i I_i  + c_Z^i Z_i + c_{2}^{i}\ket{2_i}\bra{2_i},
\label{eq:ro_model_oneq}
\end{equation}
also with real-valued $c_I^i$ and $c_Z^i\in [-1,1]$.

When correlating measurements on transmons $i$ and $j$, we compute the product of the 1-bit discretized output for each transmon.
The expected value of $m_{ji}= m_j \times m_i$ is given by $\langle m_{ji}\rangle =\textrm{Tr}\left(M_{ji} \rhoexp\right)$, where the measurement operator (also in view of the low crosstalk) is modelled as
\begin{equation}
M_{ji} = \sum_{k,l} c_{lk}^{ji}\ket{l_j k_i}\bra{l_j k_i},
\label{eq:ro_model_twoq}
\end{equation}
with real-valued coefficients $c_{lk}^{ji} \in [-1,1]$.
Making use of the 2-qubit Pauli operators given by tensor products, and again truncating to three transmon levels, we can rewrite the measurement operator in the form
%\[
%M_{ji} = c_{II}^{ij} I_j I_i  + c_{IZ}^{ij} I_j Z_i  + c_{ZI}^{ij} Z_j I_i  + c_{ZZ}^{ij} Z_j Z_i  + \sum_{k \vee l \geq 2} c_{lk}^{ji}\ket{l_j k_i}\bra{l_j k_i}.
%\]
\[
\begin{split}
M_{ji}=& c_{II}^{ji} I_j I_i  + c_{IZ}^{ji} I_j Z_i  + c_{ZI}^{ji} Z_j I_i  + c_{ZZ}^{ji} Z_j Z_i  \\
&+ c_{2I}^{ji} \ket{2_j}\!\bra{2_j}I_i + c_{2Z}^{ji} \ket{2_j}\!\bra{2_j}Z_i+ c_{Z2}^{ji} I_j\ket{2_i}\!\bra{2_i} + c_{Z2}^{ji} Z_j\ket{2_i}\!\bra{2_i}+ c_{22}^{ji}\ket{2_j 2_i}\!\bra{2_j 2_i}.
\end{split}
\]

In experiment, we calibrate the coefficients  $c_P^i$ and $c_{PQ}^{ji}$ $(P,Q\in{I,Z})$ by linear inversion of the experimental average of single-transmon and correlation measurements with the four transmons prepared in each of the 16 computational states (for which $\langle P_i\rangle=\pm 1$ and $\langle Q_j P_i\rangle=\pm 1$). We do not calibrate the coefficients $c_2^i$, $c_{2P}^{ji}$, $c_{Q2}^{ji}$, or $c_{22}^{ji}$.

Measurement pre-rotations change the measurement operator as follows: $\Rx(180\degrees)$ on transmon $i$ transforms $Z_i\rightarrow -Z_i$; $\Rx(\pm90\degrees)$ transforms $Z_i\rightarrow \pm Y_i$; and  $\Ry(\pm90\degrees)$ transforms $Z_i\rightarrow \mp X_i$. Pre-rotations do not transform the projectors $\ket{2_i}\!\bra{2_i}$ as they only act on the qubit subspace.

\subsection{Cost function evaluation}
\begin{figure*}[h!]
\begin{center}
\includegraphics[width=\textwidth]{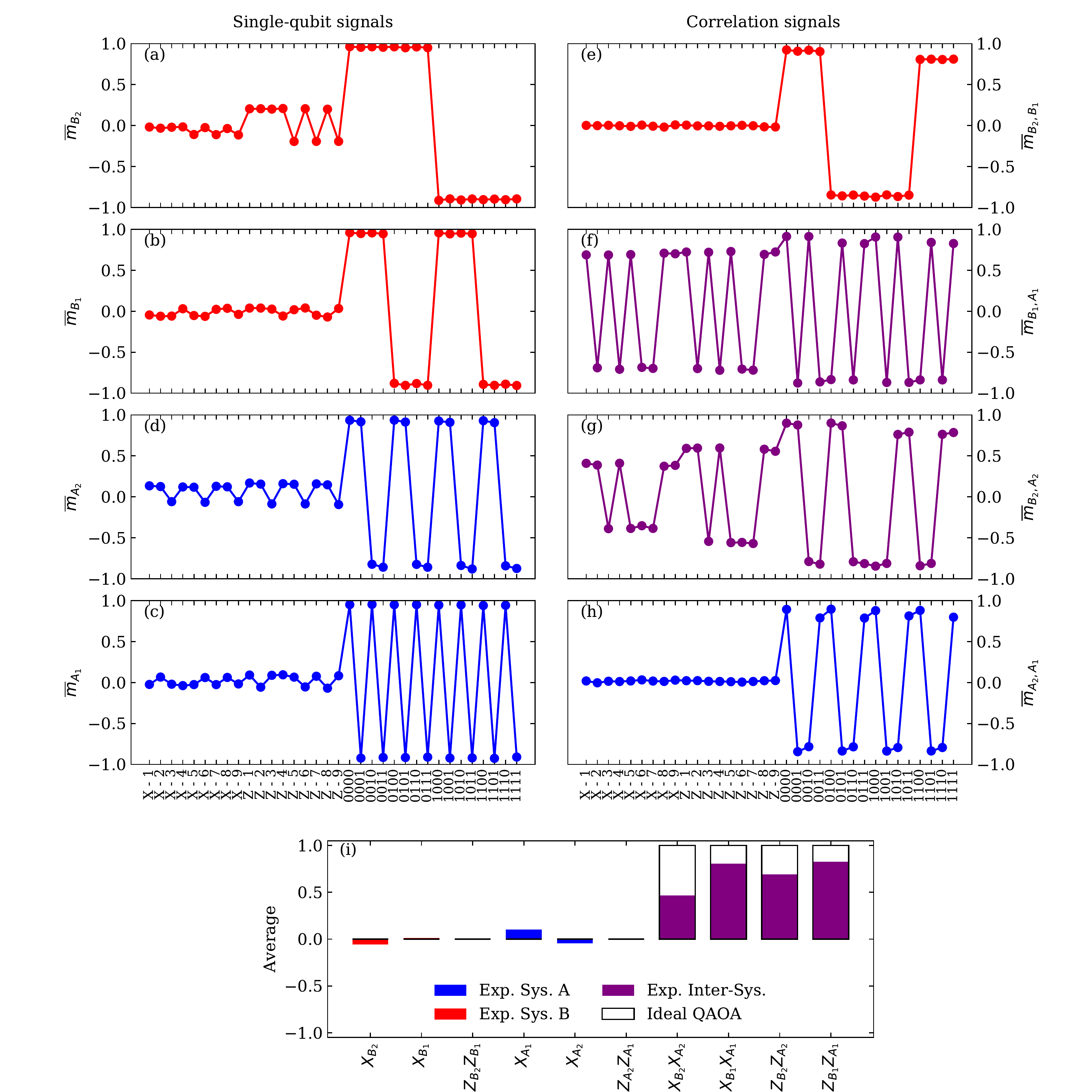}
\end{center}
\caption{
Example raw data for evaluating the cost function with variational angles $(\alphavec,\gammavec)=0$. Experimental average (from $4096$ shots) of single-qubit and correlation measurements for the 18 measurement bases in Table~S2 and for calibration of the measurement operators (following preparation of the 16 computational states of the four transmons). Panels show the experimental averages of (a-d) single-transmon measurements, (e,h) intra-system correlations and (f,g) inter-system correlations. (i) Bar graph of the experimental estimate of the expected values of all terms in $\cost$, obtained by linear inversion on the measurement averages. For comparison, we also show are the expected valuees of all the terms for an ideal processor.
}
\label{fig:SOM_raw_data}
\end{figure*}

To evaluate the cost function $\cost$, we must estimate the expected value of all single-qubit Pauli operators $X_i$, the two intra-system two-qubit Pauli operators $Z_j Z_i$, and the inter-system two-qubit Pauli operators
$X_jX_j$ and $Z_jZ_i$ [the latter only between corresponding qubits in the two systems (e.g., $\mathrm{B}_1$ and $\mathrm{A}_1$)]. We estimate these by linear inversion of the experimental averages (based on 4096 measurements) of single-transmon and relevant correlation measurements with the transmons measured in the bases specified in Table~S2. As an example, \cref{fig:SOM_raw_data}. shows the raw data for the estimation of $\cost$ with variational parameters $\left(\alphavec, \gammavec\right)=0$. Note that every evaluation of the cost function includes  readout calibration measurements to extract measurement-operator coefficients $c_P^i$ and $c_{PQ}^{ji}$ $(P,Q\in{I,Z})$.

\begin{table}[h!]
\begin{tabular}{ c | c | c }
\hline
\# & $Z$ basis measurements & $X$ basis measurements \\
\hline
1& $+\ZBtwo, +\ZBone, +\ZAtwo, +\ZAone$ & $+\XBtwo, +\XBone, +\XAtwo, +\XAone$\\
2& $+\ZBtwo, +\ZBone, +\ZAtwo, -\ZAone$ & $+\XBtwo, +\XBone, +\XAtwo, -\XAone$\\
3& $+\ZBtwo, +\ZBone, -\ZAtwo, +\ZAone$ & $+\XBtwo, +\XBone, -\XAtwo, +\XAone$\\
4& $+\ZBtwo, -\ZBone, +\ZAtwo, +\ZAone$ & $+\XBtwo, -\XBone, +\XAtwo, +\XAone$\\
5& $-\ZBtwo, +\ZBone, +\ZAtwo, +\ZAone$ & $-\XBtwo, +\XBone, +\XAtwo, +\XAone$\\
6& $+\ZBtwo, +\ZBone, -\ZAtwo, -\ZAone$ & $+\XBtwo, +\XBone, -\XAtwo, -\XAone$\\
7& $-\ZBtwo, -\ZBone, +\ZAtwo, +\ZAone$ & $-\XBtwo, -\XBone, +\XAtwo, +\XAone$\\
8& $+\ZBtwo, -\ZBone, +\ZAtwo, -\ZAone$ & $+\XBtwo, -\XBone, +\XAtwo, -\XAone$\\
9& $-\ZBtwo, +\ZBone, -\ZAtwo, +\ZAone$ & $-\XBtwo, +\XBone, -\XAtwo, +\XAone$\\
\end{tabular}
\caption{
List of the basis measurements used to evaluate all the terms in the cost function $\cost$. The bases in the left column are used to estimate intra- and inter-system terms $\langle Z_j Z_i \rangle$, while those on the right are used to extract single-qubit terms $\langle X_i\rangle$ and inter-system terms $\langle X_j X_i \rangle$.
}
\label{table:prerots}
\end{table}

\subsection{Two-qubit state tomography}
After optimization, we perform two-qubit state tomography of each system separately to assess peformance. To do this, we obtain experimental averages (from $16394$ shots) of single-transmon and correlation measurements using an over-complete set of measurement bases. This set consists of the 36 bases obtained by using all combinations of bases for each transmon $i$ and $j$, drawn from the set $\{+X,+Y,+Z,-X, -Y, -Z\}$. The expectation values of single-qubit and two-qubit Pauli operators are estimated by least-squares linear inversion. Finally, these are used to construct
\[
\rhoexp
=\frac{1}{4} \left(I_j I_i
+ \sum_{P\in{X,Y,Z}}\langle P_i \rangle_{\mathrm{est}}       I_j P_i
+ \sum_{P\in{X,Y,Z}}\langle P_j \rangle_{\mathrm{est}}       P_j I_i
+ \sum_{P,Q\in{X,Y,Z}}\langle Q_j P_i \rangle_{\mathrm{est}} Q_j P_i\right)
\]

\section{Impact of leakage two-qubit tomography}
\label{sec:eight}
Our linear inversion procedure for converting measurement averages into estimates of the expected value of one- and two-qubit Pauli operators is only valid for $\langle \ket{l_j k_i}\bra{l_j k_l}\rangle=0$ whenever either $k$ or $l\geq2$, i.e., there is no leakage on either transmon. It is therefore essential, particularly for simulation model 4, to understand precisely how leakage in either or both transmons infiltrates our extraction of the two-qubit density matrix $\rhoexp$.

First we consider the estimation of expected values of single-qubit Pauli operators  $\langle P_i \rangle$, taking$\langle Z_i \rangle$ as a concrete example.
The expected value of all measurements on this transmon for the basis combinations (6 in total) where this specific transmon is measured in the $+Z$ basis is:
\[
\langle m_{i}^{(+)} \rangle = c_I^i \langle I_i \rangle + c_Z^i \langle Z_i \rangle + c_2^i \langle\ket{2_i}\!\bra{2_i}\rangle.
\]
In turn, the expected value of all measurements on this transmon for the basis combinations (6 in total) where this specific transmon is measured in the $-Z$ basis is:
\[
\langle m_{i}^{(-)} \rangle = c_I^i \langle I_i \rangle - c_Z^i \langle Z_i \rangle + c_2^i \langle\ket{2_i}\!\bra{2_i}\rangle.
\]
Note that the contribution from the leakage term is unchanged because the pre-rotation ($I_i$ or $\Rx(180\degrees)$) only acts on the qubit subspace.
Our least-squares linear inversion of these 12 experimental averages to estimate $\langle Z_i \rangle$ is
\[
\langle Z_i \rangle_{\mathrm{est}}=\frac{1}{12 c_Z^i} \left( \sum_{+} \overline{m}_{i}^{(+)} - \sum_{-} \overline{m}_{i}^{(-)} \right).
\]
Clearly, owing to the balanced nature of this linear combination (all coefficients of equal magnitude, 6 positive and 6 negative), this estimator is not biased by $c_2^i$. In other words,
the average of $\langle Z_i \rangle_{\mathrm{est}}$ is independent of the value of $c_2^i$.

Consider now the estimation of the expected value of two-qubit Pauli operators $\langle Q_j P_i \rangle$, taking $\langle X_j Z_i \rangle$ as a concrete example.
There are four correlation measurements that contain this term.
For measurement bases $+X_j$ and $+Z_i$,
\[
\begin{split}
\langle m_{ji}^{(++)} \rangle =& c_{II}^{ji} \langle I_j I_i \rangle + c_{IZ}^{ji}\langle I_j Z_i \rangle + c_{ZI}^{ji}\langle X_j I_i \rangle + c_{ZZ}^{ji}\langle X_j Z_i \rangle\\
&+c_{II}^{ji} \langle \ket{2_j}\!\bra{2_j} I_i \rangle + c_{2Z^i}^{ji} \langle \ket{2_j}\!\bra{2_j} Z_i \rangle +c_{I2}^{ji} \langle  I_j \ket{2_i}\!\bra{2_i} \rangle + c_{Z2}^{ji} \langle X_j \ket{2_i}\!\bra{2_i} \rangle + c_{22}^{ji} \langle \ket{2_j 2_i}\!\bra{2_j 2_i} \rangle
\end{split}
\].

For measurement bases $+X_j$ and $-Z_i$,
\[
\begin{split}
\langle m_{ji}^{(+-)} \rangle =& c_{II}^{ji} \langle I_j I_i \rangle - c_{IZ}^{ji}\langle I_j Z_i \rangle + c_{ZI}^{ji}\langle X_j I_i \rangle - c_{ZZ}^{ji}\langle X_j Z_i \rangle\\
&+c_{II}^{ji} \langle \ket{2_j}\!\bra{2_j} I_i \rangle - c_{2Z^i}^{ji} \langle \ket{2_j}\!\bra{2_j} Z_i \rangle +c_{I2}^{ji} \langle  I_j \ket{2_i}\!\bra{2_i} \rangle + c_{Z2}^{ji} \langle X_j \ket{2_i}\!\bra{2_i} \rangle + c_{22}^{ji} \langle \ket{2_j 2_i}\!\bra{2_j 2_i} \rangle
\end{split}
\].

For measurement bases $-X_j$ and $+Z_i$,
\[
\begin{split}
\langle m_{ji}^{(-+)} \rangle =& c_{II}^{ji} \langle I_j I_i \rangle + c_{IZ}^{ji}\langle I_j Z_i \rangle - c_{ZI}^{ji}\langle X_j I_i \rangle - c_{ZZ}^{ji}\langle X_j Z_i \rangle\\
&+c_{II}^{ji} \langle \ket{2_j}\!\bra{2_j} I_i \rangle + c_{2Z^i}^{ji} \langle \ket{2_j}\!\bra{2_j} Z_i \rangle +c_{I2}^{ji} \langle  I_j \ket{2_i}\!\bra{2_i} \rangle - c_{Z2}^{ji} \langle X_j \ket{2_i}\!\bra{2_i} \rangle + c_{22}^{ji} \langle \ket{2_j 2_i}\!\bra{2_j 2_i} \rangle
\end{split}
\].

Finally, for measurement bases $-X_j$ and $-Z_i$,
\[
\begin{split}
\langle m_{ji}^{(--)} \rangle =& c_{II}^{ji} \langle I_j I_i \rangle - c_{IZ}^{ji}\langle I_j Z_i \rangle - c_{ZI}^{ji}\langle X_j I_i \rangle + c_{ZZ}^{ji}\langle X_j Z_i \rangle\\
&+c_{II}^{ji} \langle \ket{2_j}\!\bra{2_j} I_i \rangle - c_{2Z^i}^{ji} \langle \ket{2_j}\!\bra{2_j} Z_i \rangle +c_{I2}^{ji} \langle  I_j \ket{2_i}\!\bra{2_i} \rangle - c_{Z2}^{ji} \langle X_j \ket{2_i}\!\bra{2_i} \rangle + c_{22}^{ji} \langle \ket{2_j 2_i}\!\bra{2_j 2_i} \rangle
\end{split}
\].

Our least-squares linear inversion of these 4 experimental averages to estimate $\langle X_j Z_i \rangle$ is
\[
\langle X_j Z_i \rangle_{\mathrm{est}}=\frac{1}{4 c_{ZZ}^{ji}} \left( \overline{m}_{ji}^{(++)} - \overline{m}_{ji}^{(+-)} - \overline{m}_{ji}^{(-+)} + \overline{m}_{ji}^{(--)} \right).
\]
Clearly, owing to the balanced nature of this linear combination (all coefficients of equal magnitude, 2 positive and 2 negative), this estimator is not biased by
$c_{2I}^{ji}$, $c_{2Z}^{ji}$, $c_{I2}^{ji}$, $c_{Z2}^{ji}$, and $c_{22}^{ji}$. In other words, the average of $\langle X_j Z_i \rangle_{\mathrm{est}}$ is independent of the value of
these coefficients.

\begin{figure}[h!]
 \begin{center}
   \includegraphics[width=0.9\textwidth]{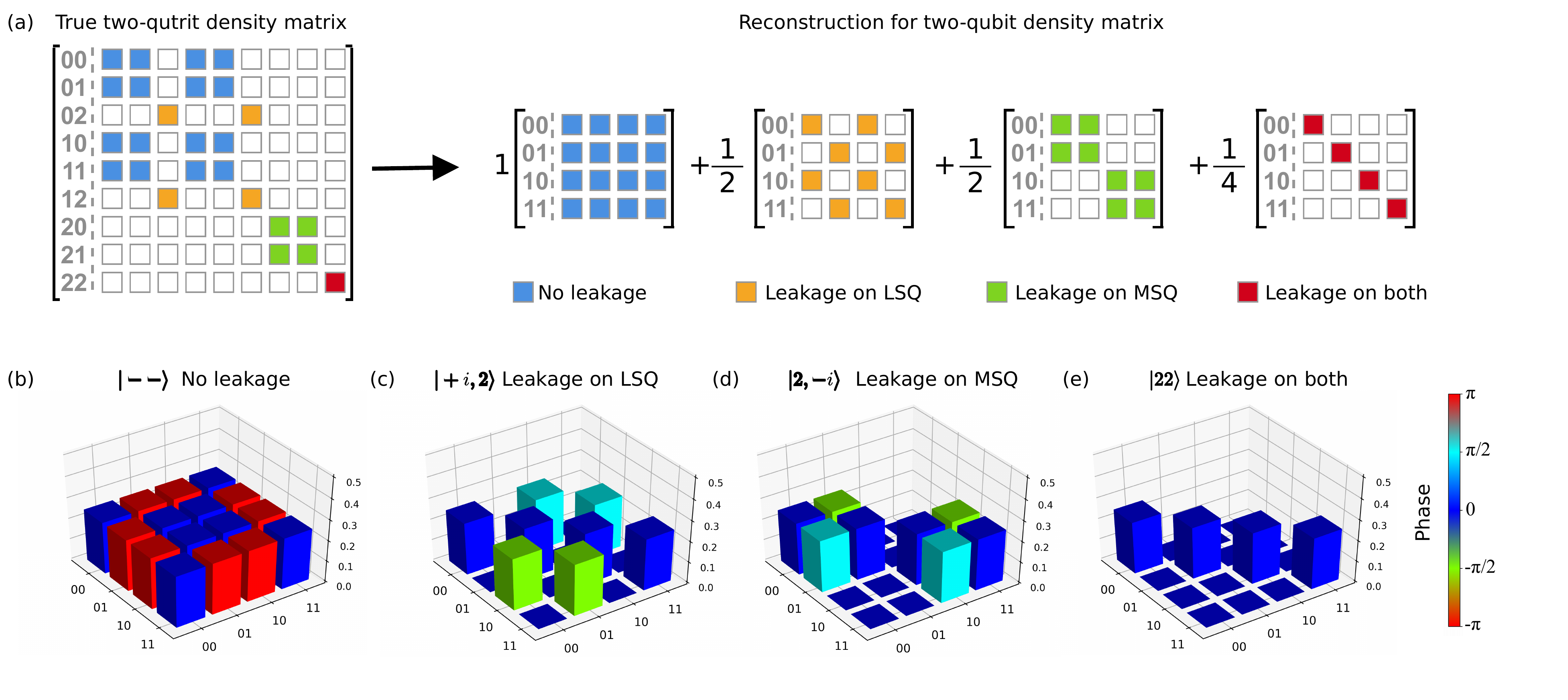}
 \end{center}
\caption{
Illustration of the impact of leakage on two-qubit state tomography.
(a) Mapping of  two-qutrit density matrix elements onto elements of the two-qubit density matrix $\rhoexp$ returned by our tomographic reconstruction procedure.
Elements in the qubit subspace (blue) are mapped correctly. Elements corresponding to one transmon in the leaked state (orange and green) are mapped onto a two-qubit state with the corresponding qubit fully mixed.
The state with leakage on both transmons (red) is mapped to the fully mixed two-qubit state.  (b-e) Example mappings for the states (b) $\ket{--}$, (c) $\ket{+i,2}$, (d) $\ket{2,-i}$, and (e) $\ket{22}$, where
$\ket{-}=\frac{1}{\sqrt{2}}\left(\ket{0}-\ket{1}\right)$ and
$\ket{\pm i}=\frac{1}{\sqrt{2}}\left(\ket{0}\pm i \ket{1}\right)$.
}
\label{fig:SOM_leak_tomo}
\end{figure}

We are finally in position to describe how leakage in the two-transmon system infiltrates into our two-qubit tomographic reconstruction procedure.
Evidently, the complete description of the two-transmon system would be a two-qutrit density matrix $\rhotwoqutrit$, but our procedure returns a two-qubit density matrix $\rhoexp$.
It is therefore key to understand how elements of $\rhotwoqutrit$ are mapped onto $\rhoexp$. Table~S3 summarizes these mappings and \cref{fig:SOM_leak_tomo} illustrates them, including several examples.
We have verified the mappings by exactly replicating the tomographic procedure in our numerical simulation using \textit{quantumsim}. To incorporate this into the simulations, we have made use of the measurement coefficients $c_i$ experimentally obtained. These leakage mappings have also been used when adding leakage in simulation model 4.

\begin{table}[h!]
\begin{tabular}{ c | c  }
\hline
Elements of $\rhotwoqutrit$ & Mapping onto $\rhoexp$\\
\hline
\hline
&\\
$\ket{l'_j,k'_i}\bra{l_j,k_i}$        &       $\ket{l'_j,k'_i}\bra{l_j,k_i}$            \\
&\\
$\ket{l'_j,2_i}\bra{l_j,2_i}$        &       $\frac{1}{2}\ket{l'_j}\bra{l_j} I_i$     \\
&\\
$\ket{2_j,k'_i}\bra{2_j,k_i}$        &       $\frac{1}{2} I_j \ket{k'_i}\bra{k_i}$    \\
&\\
$\ket{2_j,2_i}\bra{2_j,2_i}$        &       $\frac{1}{4} I_j I_i$                   \\
&\\
All other elements                  &       $0$                                     \\
\end{tabular}
\caption{
Mapping of the elements of $\rhotwoqutrit$ to $\rhoexp$ by our two-qubit state tomography procedure. Here, $k,l,k',l'\in\{0,1\}$ denote computational (unleaked) states.
}
\label{table:mappings}
\end{table}

\section{Error model for numerical simulations}
\label{sec:nine}

Our numerical simulations use the \textit{quantumsim}~\cite{Quantumsim} density-matrix simulator with the error model described in the \textit{quantumsim\_dclab} subpackage.

Single-qubit gates are modeled as perfect rotations, sandwiched by two $10~\ns$ idling blocks.
The idling model takes into account amplitude damping ($\Tone$), phase damping ($\Ttwoecho$) (noise model 1 of the main text), and residual $ZZ$ crosstalk (noise model 3).
To implement it, we first split the idling intervals into slices of $10~\ns$ or less. These slides include amplitude and phase damping. Between these slices, we add instantaneous two-qubit gates capturing the residual coupling described by the Hamiltonian:
\begin{equation}
  H = \zeta_{ij} \ket{11} \bra{11} = - \frac{\zeta_{ij}}{2} \left(1 - Z_i -Z_j + Z_i Z_j \right).
\end{equation}
The measurements of $\Tone$, $\Ttwoecho$ and $\zeta_{ij}$ are detailed in previous sections.

The error model for CZ gates is described in detail in~\cite{Varbanov20}.
The dominant error sources are identified to be leakage of the fluxed transmon to the second-excited state through the $\ket{11} \leftrightarrow \ket{02}$ channel and increased dephasing (reduced $\Ttwoecho$) due to the fact that the fluxed transmon is pulsed away from the sweetspot. These two effects implement noise models 4 and 2, respectively.
The two-transmon process is modeled as instantaneous, and sandwiched by two idling blocks of $35~\ns$ with decreased $\Ttwoecho$ on the fluxed transmon.
Quasistatic flux noise is suppressed to first order in the Net Zero scheme and is therefore neglected.
Residual $ZZ$ crosstalk is not inserted during idling for the transmon pair, because it is absorbed by the gate calibration.
The error model described in~\cite{Varbanov20} allows for higher-order leakage effects, e.g., so-called leakage conditional phases and leakage mobility. We do not include these effects.

The simulation finishes by including the effect of leakage on the tomographic procedure as discussed in \cref{sec:eight}. The density matrix is obtained at the qutrit level, and the correct mapping for the density-matrix elements is applied. We take special care to use the experimental readout coefficients $c_i$ to model the readout signal for the simulated density matrix, according to \cref{eq:ro_model_oneq} and \cref{eq:ro_model_twoq}. The simulation produces data for the same basis set as shown in \cref{fig:SOM_raw_data}. Afterwards, the same tomographic state reconstruction routine as in the experiment is applied to these data. in this way, noise model 4 properly accounts for the imperfect reconstruction of leaked states, providing a fair comparison to experiment.

%%%%%%%%% REFERENCES 2 %%%%%%%%%
\end{bibunit}
\end{document}